\documentclass[aps,prb,twocolumn,showpacs,citeautoscript,superscriptaddress,longbibliography]{revtex4-1}

\usepackage{graphicx}  % needed for figures
\usepackage{epstopdf}
\usepackage{dcolumn}   % needed for some tables
\usepackage{bm}        % for math
\usepackage{amsmath}
\usepackage{amssymb}   % for math
\usepackage{wrapfig}
\usepackage{array}
\usepackage[caption=false]{subfig}
\usepackage[bookmarks=false,linkcolor=blue,urlcolor=blue,colorlinks,citecolor=blue]{hyperref}
\usepackage{exscale,relsize}
\usepackage{color}
\usepackage{times, verbatim}
\usepackage{bbold}

\newcommand{\ket}[1]{|#1\rangle}

\makeatletter

\begin{document}

\title{Number-conserving analysis of measurement-based braiding with Majorana zero modes }

\author{Christina Knapp}
\affiliation{Department of Physics, University of California, Santa Barbara,
	California 93106 USA}
\affiliation{Department of Physics and Institute for Quantum Information and Matter, California Institute of Technology, Pasadena,
	California 91125 USA}
\affiliation{Walter Burke Institute for Theoretical Physics, California Institute of Technology, Pasadena,
	California 91125 USA}

\author{Jukka I. V\"ayrynen}
\affiliation{Station Q, Microsoft Corporation, Santa Barbara, California 93106-6105 USA}

\author{Roman M. Lutchyn}
\affiliation{Station Q, Microsoft Corporation, Santa Barbara, California 93106-6105 USA}

\begin{abstract}
Majorana-based quantum computation seeks to encode information non-locally in pairs of Majorana zero modes (MZMs), thereby isolating qubit states from a local noisy environment.  In addition to long coherence times, the attractiveness of Majorana-based quantum computing relies on achieving topologically protected Clifford gates from braiding operations. Recent works have conjectured that mean-field BCS calculations may fail to account for non-universal corrections to the Majorana braiding operations. Such errors would be detrimental to Majorana-based topological quantum computing schemes. In this work, we develop a particle-number-conserving approach for measurement-based topological quantum computing and investigate the effect of quantum phase fluctuations. We demonstrate that braiding transformations are indeed topologically protected in charge-protected Majorana-based quantum computing schemes.

\end{abstract}

\date{\today}

\maketitle

%%%
\section{Introduction}

Topological quantum computation is predicated on the idea that information stored non-locally in pairs of non-Abelian anyons or topological defects is robust to local noise sources~\cite{Nayak08, DasSarma2015}.  Braiding the anyons or defects implements a non-trivial operation on the quantum state, while preserving the topological protection of the encoded information.  Topological protection is generally defined as exponentially suppressed scaling of error rates in parameter ratios of the system that can be made large.

At present, the most promising approach towards realizing topological quantum computing utilizes Majorana zero modes (MZMs), non-Abelian topological defects of a superconductor~\cite{Kitaev01,Kitaev03,Alicea12a,Lutchyn17}.  Each MZM is described by a Majorana operator, $\gamma_j=\gamma_j^\dagger$, satisfying anticommutation relations
\begin{align}\label{eq:Maj-com}
\{\gamma_j,\gamma_k\} &= 2\delta_{j,k}.
\end{align}
Majorana-based qubits encode quantum information in the fermion parity of pairs of MZMs, corresponding to the operator $i\gamma_j\gamma_k$.  Braiding MZMs $j$ and $k$ corresponds to the operator~\cite{Read2000, Ivanov01}
\begin{align}\label{eq:braid}
R^{(jk)} &= \frac{1+\gamma_j\gamma_k}{\sqrt{2}}.
\end{align}
Braiding, combined with a two-qubit entangling measurement, is sufficient to implement all Clifford operations.  Supplementing braiding and measurement with a non-Clifford gate ({\it e.g.}, using magic state distillation, which also benefits from protected Clifford gates) enables universal quantum computation~\cite{Nayak08, DasSarma2015}.  The attractiveness of Majorana-based quantum computing is equally dependent on achieving long coherence times for the idle qubit, and on achieving topologically protected Clifford operations.

There has been impressive experimental progress in tuning semiconductor-superconductor nanowires into a topological superconducting phase hosting MZMs at either endpoint~\cite{Lutchyn10,Oreg10,Mourik12,Deng12,Das12,Churchill13,Finck12,Deng16,Albrecht16,Nichele17,Zhang17,Vaitiekenas18}.  The continued experimental improvement of these systems has led to theoretical interest in designing Majorana-based qubits out of such heterostructures~\cite{Sau2010a, Hassler11,vanHeck11,Hyart13,Aasen16}.  In particular, several works in the last few years have proposed charge-protected Majorana-based qubits~\cite{Karzig17,Plugge17,Vijay16}.  These qubits have a large charging energy to suppress extrinsic quasiparticle poisoning ({\it i.e.,} stochastic electron tunneling into a Majorana island that changes the topological state of the system).  Additionally, these qubits are operated according to a measurement-based braiding protocol~\cite{Bonderson08b,Bonderson08c,Zhang16,Karzig17,Vijay16, Plugge17,Tran19,Bomantara19} to circumvent the difficulty of physically moving MZMs in 1D wire networks~\cite{Alicea11,Bauer18} and the susceptibility of anyon braiding to problematic diabatic errors~\cite{Knapp16}. 

Charge-protected Majorana-based qubits are operated in the Coulomb-blockaded regime, for which quantum phase fluctuations of the superconducting order parameter are important. The majority of previous studies of Majorana systems have used mean-field BCS models, which do not take into account such fluctuations. A natural question to consider is the extent to which mean-field results apply to a physical system with particle-number conservation~\cite{Ortiz14,Ortiz16,Wang17,Wang18,Lin17,Lin18}.  Field-theoretic bosonization has emerged as a useful tool for comparing mean field and number-conserving predictions for 1D topological superconductors~\cite{Sau11,Fidkowski11,Cheng15,Knapp17}.  
Previous works have demonstrated that Majorana nanowires have a topologically protected ground state degeneracy even in the absence of long-range superconducting order~\cite{Fidkowski11}, examined the fractional Josephson effect in Coulomb-blockaded Majorana-based devices~\cite{Cheng15}, calculated the charge distribution associated with the topological state and thus the susceptibility of Majorana-based qubits to noise~\cite{Knapp17}.  These studies have reaffirmed the topological protection of an idle charge-protected Majorana-based qubit.

Majorana-based quantum computation additionally relies on MZM braiding to implement topologically protected Clifford gates.  Recent studies have questioned whether number conservation introduces non-universal corrections to the Majorana braiding transformations.  References~\onlinecite{Lin17,Lin18} used number projected Bogoliubov-de-Gennes theory for 2D p+ip superconductors to argue that Cooper pair coupling to local observables may affect MZM braiding in 2D p+ip superconductors.  The potential braiding phase errors raised by Refs.~\onlinecite{Lin17,Lin18} would be detrimental to the field of Majorana-based quantum computing and thus warrant serious investigation.

In this work, we extend the bosonized formalism of Refs.~\onlinecite{Fidkowski11, Cheng15,Knapp17} to study measurement-based braiding for charge-protected Majorana-based qubits.  In particular, we examine whether the MZM parity measurements proposed in Ref.~\onlinecite{Karzig17} are susceptible to non-universal corrections from quantum fluctuations of the superconducting phase, and the implications for measurement-based braiding.  We find:

\begin{enumerate}

\item In the absence of charging energy, the left/right end of the proximitized wire segment $j$ hosts a charged fermionic zero mode $\Gamma_{j,L/R}$. The neutral product of two such operators $i\Gamma_{j,J}^\dagger\Gamma_{k,K}$, ${J,K\in\{L/R\}}$, is closely related to the MZM parity.

\item The quantum dot-based tunneling measurement proposed in Ref.~\onlinecite{Karzig17} couples to the MZM parity.  Corrections to this measurement from number conservation occur outside of the ground state subspace and are therefore exponentially suppressed in the charge gap over the temperature.  Spatial quantum phase fluctuations in the superconductor reduce the measurement visibility, but do not otherwise affect projective parity measurements.

\item The quantum dot-based tunneling measurement can be used in a measurement-based braiding protocol.  As quantum fluctuations in the superconductor do not preclude projective measurements, the operation implemented by this protocol simulates a topologically protected braiding transformation.
\end{enumerate}

The remainder of this paper is organized as follows. In Sec.~\ref{sec:setup}, we describe our model of the charge-protected qubit displayed in Fig.~\ref{fig:setup}.  We then derive the zero modes at each end of the proximitized segments and demonstrate their anticommutation as well as other key properties, see Sec.~\ref{sec:zm}.  We  identify the MZM parity and demonstrate that it is insensitive to all local operators, up to exponentially suppressed terms. In Sec.~\ref{sec:mst}, we then consider the quantum dot-based tunneling measurement depicted in Fig.~\ref{fig:setup}.  We show that such a measurement couples to the MZM parity.  Finally, in Sec.~\ref{sec:braiding} we argue that the measurement-based braiding protocol outlined in Ref.~\onlinecite{Karzig17} is topologically protected.  We conclude by identifying the role number conservation plays throughout our analysis and discussing the connection to previous works in Secs.~\ref{sec:comparison} and \ref{sec:conclusions}.  We relegate technical details of the calculations to the appendices.

%%%
\section{Setup}\label{sec:setup}

\begin{figure*}[t]
\begin{center}
	\includegraphics[width=2\columnwidth]{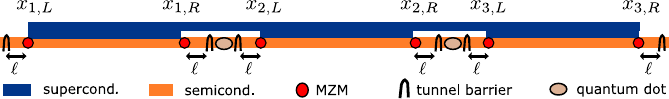}
	\caption{
Basic qubit layout proposed in Ref.~\onlinecite{Karzig17}.  A semiconductor (orange) is proximitized by a superconductor (blue) in three spatial regions, $x_{j,L}<x<x_{j,R}$ for $j\in\{1,2,3\}$ and $L/R$ indicating left/right.  At the end of each proximitized segment, there is a bare semiconductor region of length $\ell$, terminated by a tunnel barrier.  Each bare semiconductor region hosts a charged fermionic zero mode $\Gamma_{j,J}$, where the neutral product $i\Gamma_{j,J}^\dagger \Gamma_{k,K}$ corresponds to the MZM parity $i\gamma_{j,J}\gamma_{k,K}$.  The regions between two proximitized wires hosts a quantum dot.  To perform a measurement, the barriers are lowered to permit tunneling between the quantum dot and the bare semiconducting regions.  Reference~\onlinecite{Karzig17} discusses how the same physics can be used to measure any pair of MZMs using coherent links (floating topological superconductors in a fixed fermion parity state).  Our analysis generalizes straightforwardly to the non-linear qubit structures proposed in Ref.~\onlinecite{Karzig17}.
}
	\label{fig:setup}
\end{center}
\end{figure*}

We consider the charge-protected Majorana-based qubit depicted in Fig.~\ref{fig:setup}.  The full structure of the qubit will only be important in Sec.~\ref{sec:braiding} when we consider measurement-based braiding (which requires a minimum of six MZMs).  We highlight the relevant physics below.

A spinless semiconducting nanowire (orange) is proximitized by an s-wave superconductor (dark blue) in three segments.  Each segment is connected to a superconducting backbone, which is assumed to have many channels so that there is no relative charging energy between different regions.  A tunnel barrier separates the end of each proximitized region from a quantum dot or lead that can be used for a tunneling measurement, see Section~\ref{sec:mst}.  The device in Fig.~\ref{fig:setup} hosts six MZMs (red dots), one at each end of the proximitized nanowires.  We label the proximitized wires by $j\in\{1,2,3\}$ and left/right end of the wires by $J\in\{L/R\}$.  Below, we refer to the MZM at the $J$th end of the $j$th wire as $\gamma_{j,J}$.  The qubit forms a floating (non-grounded) superconducting island with four degenerate (up to exponentially suppressed corrections that we neglect here) ground states.  Two of these states constitute the computational basis, while the remaining two are ancilla degrees of freedom used to facilitate measurement-based braiding, see Section~\ref{sec:braiding}.  Our analysis of the device shown in Fig.~\ref{fig:setup} generalizes straightforwardly to the non-linear geometries proposed in Ref.~\onlinecite{Karzig17}.

 We study this device using a number-conserving bosonized formalism, previously used in Refs.~\onlinecite{Fidkowski11,Cheng15,Knapp17, Snizhko18}.  We model the semiconductor with spinless electrons defined by
\begin{align}
\psi_\text{sm}(x) &\sim e^{ik_F x} e^{i\theta(x) + i\phi(x)} + e^{-ik_F x} e^{i\theta(x)-i\phi(x)}.
\end{align}
In the above notation, $\theta$ and $\phi$ are bosonic operators whose commutator 
\begin{align}\label{eq:commutator}
[\phi(x),\theta(y)] &= i\pi \Theta(x-y),
\end{align}
ensures that electron operators at distinct points anticommute.  The charge density is related to $\phi$ by $\rho(x)=\partial_x\phi(x)/\pi$, while the operator $e^{i\theta(x)}$ adds a charge to the semiconductor at position $x$.  In the above, $k_F$ is the semiconductor Fermi momentum and $\Theta(x)$ is the Heaviside function.

The superconductor carries both charge ($\rho$) and spin ($\sigma$) fields
\begin{align}
 \psi_{\text{sc},\uparrow/\downarrow}(x) \sim& e^{ik_F^{(\rho)}x} e^{\frac{i}{\sqrt{2}}\left( \theta_\rho(x) + \phi_\rho(x)\pm \left[\theta_\sigma(x) + \phi_\sigma(x)\right]\right)} \notag
\\ &+ e^{-ik_F^{(\rho)}x} e^{\frac{i}{\sqrt{2}} \left(\theta_\rho(x) -\phi_\rho(x)\pm\left[\theta_\sigma(x)-\phi_\sigma(x)\right]\right)},
\end{align}
where $\uparrow$ ($\downarrow$) corresponds to $+$ ($-$) in the exponent.  Similarly, the commutation relations are
\begin{align}\label{eq:canonical-comm}
    [\phi_\lambda(x),\theta_{\lambda'}(y)]&= i \pi \delta_{\lambda,\lambda'}\Theta(x-y)
\end{align}
where $\lambda,\lambda'\in\{\rho,\sigma\}$.
The charge density in the superconductor is defined by ${\rho_\text{sc}(x) = \sqrt{2}\partial_x \phi_\rho(x)/\pi}$ and the current is $\sqrt{2}\partial_x\theta_\rho/\pi$.  Thus the operator $e^{i\theta_\rho(x)/\sqrt{2}}$ adds a charge to the superconductor at position $x$.  We denote the Fermi momentum in the superconductor by $k_F^{(\rho)}$.  The number operator for the combined semiconductor and superconductor is
\begin{align} \label{eq:N}
N &= N_\text{sm}+ N_\text{sc}
\\ N_\text{sm} &= \frac{1}{\pi} \sum_{j=1}^3 \left(\phi(x_{j,R}+\ell) -\phi(x_{j,L}-\ell)\right)
\\ N_\text{sc} &= \frac{\sqrt{2}}{\pi}  \left( \phi_\rho(x_{3,R})-\phi_\rho(x_{1,L}) \right).
\end{align}

We model the semiconductor as a Luttinger liquid and the superconductor as a  Luther-Emery liquid~\cite{Seidel05}. Due to the spin gap in the superconductor, one can integrate out spin degrees of freedom in the superconductor and obtain an effective pair tunneling Hamiltonian across the semiconductor/superconductor interface~\cite{Fidkowski11}. Thus, the effective low-energy Hamiltonian has only charge degrees of freedom, and can be written as
\begin{align}
H_\text{sm} &= \frac{v}{2\pi}\sum_{j=1}^{3} \int_{x_{j,L}-\ell}^{x_{j,R}+\ell} dx \left\{ K \left(\partial_x \theta\right)^2 + K^{-1} \left(\partial_x \phi \right)^2  \right\}
\\ H_\text{sc} &= \frac{v_\rho}{2\pi} \int_{x_{1,L}}^{x_{3,R}} dx \left\{ K_\rho \left( \partial_x \theta_\rho\right)^2 + K_\rho^{-1}\left( \partial_x \phi_\rho \right)^2 \right\}
\\ H_\text{P} &= \frac{\Delta_P}{2\pi a} \sum_{j=1}^3 \int_{x_{j,L}}^{x_{j,R}} dx \cos\left( \sqrt{2}\theta_\rho -2\theta \right)\label{eq:HP}
\end{align}
 In the above, $v$ and $K$ are the Fermi velocity and Luttinger liquid parameter for the semiconductor, while $v_\rho$ and $K_\rho$ are for the superconductor.  The term $H_\text{P}$ describes pair-tunneling between the semiconductor and superconductor.  This term is a relevant perturbation that flows to strong coupling in the infrared limit and opens up a topological superconducting gap $\Delta_P.$~\cite{Fidkowski11} As $H_\text{sm},$ $H_\text{sc}$, and $H_\text{P}$ all commute with the number operator $N$, our model is explicitly number-conserving.

When the semiconductor and superconductor are decoupled from each other, for instance in the region ${x_{j,R}<x<x_{j+1,L}}$, the semiconductor and superconductor fields introduced above are the natural degrees of freedom to describe the system.  In the $j$th proximitized wire, the pairing term in Eq.~\eqref{eq:HP} strongly couples the semiconductor and superconductor.  In this case, the convenient fields to use are
\begin{align}\label{eq:theta-}
\theta_-(x)&=\frac{1}{\sqrt{2}}\theta_\rho(x)-\theta(x)
\\ \theta_+(x)&=\frac{1}{2} \left(\frac{1}{\sqrt{2}}\theta_\rho(x) +\theta(x)\right),
\end{align}
and their respective dual fields
\begin{align}
    \phi_-(x) &= \frac{1}{2} \left(\sqrt{2}\phi_\rho(x) - \phi(x) \right)
    \\ \phi_+(x) &= \sqrt{2}\phi_\rho(x) +\phi(x).
\end{align}
Note that the total charge of a proximitized wire can be written in terms of $\phi_+$
\begin{align}
    N^j_+ &= \frac{1}{\pi} \int_{x_{j,L}}^{x_{j,R}} dx \, \partial_x \left(\sqrt{2}\phi_\rho +\phi\right)
    \\ &= \frac{1}{\pi} \int_{x_{j,L}}^{x_{j,R}} dx \, \partial_x \phi_+
\end{align}
 and commutes with $\theta_-$. Henceforth, we will derive an effective low-energy theory for the system. At energies $\varepsilon \ll \Delta_P$, the field $\theta_-(x)$ for each proximitized wire is pinned and takes values $\theta_-=0$ or $\pi.$  The even and odd superpositions of these minima,
\begin{align}\label{eq:parity-eigenstates}
    \ket{\pm}_j &= \frac{1}{\sqrt{2}}\left(\ket{\theta_-=0}_j\pm \ket{\theta_-=\pi}_j\right)
\end{align}
are eigenstates of the relative fermion parity $(-1)^{N^j_-},$ where
\begin{align}
    N^j_- &= \frac{1}{\pi}\int_{x_{j,L}}^{x_{j,R}} dx \, \partial_x \phi_-.
\end{align}
When the total charge of the qubit is fixed, say, to be even, there are four such states: $\ket{\pm}_1\ket{\pm}_2\ket{+}_3$, and $\ket{\pm}_1\ket{\mp}_2\ket{-}_3$ where the subscript here refers to a particular proximitized segment in Fig.\ref{fig:setup}.
References~\onlinecite{Fidkowski11,Knapp17} argued that these states are indistinguishable by all local operators, have an exponentially suppressed degeneracy splitting, and are predicted to have exceptionally long coherence times.  Thus, the topological information is completely encoded in $\theta_-$.  We now extend this analysis to consider qubit measurement, with the aim of understanding whether topological protection extends to Clifford gates implemented by measurement-based braiding of MZMs.

We introduce two new elements: (1) bare semiconducting regions at the end of each proximitized wire, terminated by a tunneling barrier of potential $V_B$
\begin{align}\label{eq:HB}
H_\text{B} &= V_B \sum_{j=1}^3 \left\{ \cos\left(2\phi\left(x_{j,L}-\ell\right) \right) + \cos\left(2\phi\left(x_{j,R}+\ell \right) \right) \right\};
\end{align}
 and (2) a Hamiltonian $H_C$ describing the charging of the island
\begin{align}\label{eq:HC}
H_C &= E_C \left( N-N_g\right)^2,
\end{align}
where $N$ is defined by Eq.~\eqref{eq:N} and $N_g$ is a dimensionless gate voltage.  When operated at a Coulomb valley, {\it i.e.}, $N_g\in \mathbb{Z}$, adding or removing an electron from the island costs an energy $E_C$.  In the limit $E_C$ is much larger than the temperature $T$, single electron processes are exponentially suppressed.  This is the sense in which the qubit is ``charge-protected." Henceforth, we assume that the level spacings for the superconductor $\delta_\text{sc}$ and the semiconductor $\delta_\text{sm}$ are negligibly small. The latter applies to a sufficiently long wire, $v/L_\text{wire}\ll T$, as well as when there is a strong coupling between the superconductor and semiconductor which further suppresses $\delta_\text{sm}$ due to small $\delta_\text{sc}$~\cite{Stanescu2011}

In the remainder of the paper, we study the weak tunneling limit for the qubit-dot coupling and assume that the barrier potential $V_B$ is sufficiently large that $\phi(x_{j,L/R})$ are pinned to $m_{j,L/R}\pi$ ($m_{j,L/R}\in \mathbb{Z}$). At low energies, the pairing amplitude $\Delta_P$ pins the difference field $\theta_-$ to $n_j \pi$ for $x_{j,L}<x<x_{j,R}$ ($n_j \in \mathbb{Z}$). Finally, we assume that the superconducting field $\theta_\rho$ is spatially homogeneous throughout the superconductor due to a large number of transverse channels ({\it i.e.}, $K_\rho\to \infty$).  This constraint will be relaxed in Section~\ref{sec:comparison}.

Given the above assumptions and ${T\ll\text{min}\left(V_B,\Delta_P\right)}$, one can derive the low-energy theory by imposing mixed boundary conditions for the bare semiconducting regions at the ends of each proximitized segment.  We show below that this results in a fermionic zero mode localized in each of these regions.

%%%
\section{Zero mode solution}\label{sec:zm}

In this section, we show that the bare semiconductor region at the end of a proximitized wire localizes a fermionic zero mode.  This zero mode arises from the mixed boundary conditions in the segment - normal boundary conditions at one end ($\psi_{\text{sm}, R}=\psi_{\text{sm}, L}$ corresponding to $\phi$-field being pinned by the barrier Hamiltonian in Eq.~\eqref{eq:HB}), and Andreev boundary conditions  at the opposite end ($\psi_{\text{sm}, R}=\psi_{\text{sm}, L}^\dag$ corresponding to $\theta_-$ being pinned by the pairing term in Eq.~\eqref{eq:HP})~\cite{Fidkowski12}.

The fields in the bare semiconductor region to the $J$th side of the $j$th proximitized segment admit normal mode expansions
\begin{align}\label{eq:phi}
\phi_{j,J}(y)& \!= \phi_{j,J}^0\! + \!i\sqrt{2K}\sum_{k=0}^\infty \frac{\cos\! \left( [2k+ 1] \frac{\pi y}{2\ell}\right)} {\sqrt{2k+ 1}}  \! \left( b_k^\dagger - b_k \right)
\\  \theta_{j,J}(y) & \! = \theta_{j,J}^0 \!+\! \sqrt{\frac{2}{K}}  \sum_{k=0}^\infty  \frac{\sin\! \left( [2k+1]\frac{\pi y}{2\ell} \right)}{\sqrt{2k+1}} \! \!\left( b_k^\dagger + b_k \right) \!.\label{eq:theta}
\end{align}
The bosonic operators $b_k$ have canonical commutation relations $[b_k,b_{k'}^\dagger]=\delta_{k,k'}$, while the zero modes satisfy
\begin{align}\label{eq:zm-commutator}
    [\phi_{j,J}^0,\theta_{k,K}^0] &= i \pi \Theta(j-k+J/2),
\end{align}
where $J=L=-1$ and $J=R=+1.$
For simplicity, we have used the shifted coordinates $y=x-x_{j,J}$, which range between $[-\ell,0]$ for $J=L$ and $[0,\ell]$ for $J=R$.  One can show that the expansions in Eqs.~(\ref{eq:phi}-\ref{eq:theta}) satisfy the commutator of Eq.~\eqref{eq:commutator}, see Appendix~\ref{app:zero-mode} for details.

Equations~(\ref{eq:phi}-\ref{eq:theta}) diagonalize $H_\text{bare}$:
\begin{align}
H_\text{bare}&= J\frac{v}{2\pi}\int_0^{J\ell}  dy \left\{ K \left( \partial_{y} \theta_{j,J}\right)^2 + K^{-1} \left(\partial_{y}\phi_{j,J} \right)^2 \right\}
\\ &= \frac{\pi v}{\ell} \sum_{k=0}^\infty \left(k+\frac{1}{2}\right) \left( b_k^\dagger b_k + \frac{1}{2} \right).
\end{align}
The quasiparticle excitations in this segment have an energy gap of $\pi v/\ell$.
The bosonic zero modes
\begin{align}
\phi_{j,J}^0&=\pi m_{j,J}, & \theta_{j,J}^0 &=  \frac{\theta_\rho(x_{j,J})}{\sqrt{2}} - \pi n_j \,, \label{eq:theta0}
\end{align}
ensure that $\phi_{j,J}(y)$ and $\theta_{j,J}(y)$  satisfy the boundary conditions imposed $H_B$ and $H_P$.  Note that $\pi n_j$ is exactly the difference field $\theta_-$ for wire $j$ defined in Eq.~\eqref{eq:theta-}, which encodes the topological state of the $j$th wire.

The bare semiconductor regions localize a zero mode of the full many body spectrum of $H_\text{bare}$ $\Gamma_{j,J}$, which when projected into the ground state subspace with no excited bosons ($\langle b_k^\dagger b_k\rangle=0$), $\Gamma_{j,J}$ takes the simple form
\begin{align}\label{eq:gs-projection}
\Gamma_{j,J}&= e^{i\theta_{j,J}^0-i\phi_{j,J}^0}.
\end{align}
Equation~\eqref{eq:gs-projection} satisfies fermionic anticommutation relations, $\{\Gamma_{j,J},\Gamma_{k,K}\}=2\delta_{j,k}\delta_{J,K}.$
The derivation and ground state projection of $\Gamma_{j,J}$ closely follows that in Ref.~\onlinecite{Clarke13}, which considered a similar problem of a quantum Hall edge subject to mixed boundary conditions.  Their result was further extended to the number-conserving case by Ref.~\onlinecite{Snizhko18}.  For this reason, we relegate further details to Appendix~\ref{app:zero-mode}.  

In addition to being a zero mode of $H_\text{bare}$, $\Gamma_{j,J}$ also commutes with $H_\text{sm}+H_\text{sc}+H_\text{P}$.  However, $\Gamma_{j,J}$ has a non-trivial commutator with the number operator $N$.  Working from Eq.~\eqref{eq:gs-projection},
\begin{align}
[N,\Gamma_{j,J}] &= \frac{1}{\pi} [\phi(x_{j,R})-\phi(x_{j,L}),\theta_{j,J}^0] i \Gamma_{j,J}
= -\Gamma_{j,J}.
\end{align}
In the above, we used the relation $[A,f(B)]=[A,B]f'(B)$ when $A$ and $B$ both commute with their commutator.  It follows that $\Gamma_{j,J}$ acquires non-trivial time-dependence from $H_C$:
\begin{align}
\frac{d \Gamma_{j,J}(t)}{dt} &= i [H_C, \Gamma_{j,J}(t)]
\\ &= i E_C [(N-N_g)^2,\Gamma_{j,J}(t)]
%\\ &= i E_C  \left( (N-N_g)[N,\Gamma_{j,J}(t)]+[N,\Gamma_{j,J}(t)](N-N_g) \right)
\\ &= -i E_C \left( 2N -2N_g +1 \right) \Gamma_{j,J}(t).
\end{align}
In imaginary time, the evolution of $\Gamma_{j,J}(\tau)$ is
\begin{align}\label{eq:Gamma-time}
\Gamma_{j,J}(\tau) &= e^{- E_C \left( 2N -2N_g +1\right) \tau} \Gamma_{j,J}(0).
\end{align}
Similar logic shows
\begin{align}\label{eq:Gamma-time-dagger}
    \Gamma_{j,J}^\dagger(\tau) &= e^{E_C \left( 2N-2N_g -1\right)\tau}\Gamma_{j,J}^\dagger(0).
\end{align}

When the qubit is tuned to a Coulomb valley, {\it e.g.}, $N_g=0$ and $\langle N\rangle=0$, we have
\begin{align}\label{eq:corr-function}
    \langle T_\tau \Gamma_{j,J}^\dagger(\tau_1)\Gamma_{k,K}(\tau_2)\rangle_C &= e^{-E_C|\tau_1-\tau_2|} \langle \Gamma_{j,J}^\dagger(0)\Gamma_{k,K}(0)\rangle,
\end{align}
where the averaging is taken over charging Hamiltonian, see Appendix~\ref{app:corr-function}.  %Note that $H_C$ does not distinguish different MZM parity states.

To evaluate the equal time correlator, we first note that the zero mode operators satisfy fermionic anticommutation relations
\begin{align}
 \{\Gamma_{j,J}^\dagger,\Gamma_{k,K}\} &= 2\delta_{j,k}\delta_{J,K}.
\end{align}
The neutral product $i\Gamma_{j,J}^\dagger \Gamma_{k,K}$ is Hermitian for ${(j,J)\neq (k,K)}$ and can be written as
\begin{align}\label{eq:neutral-product}
     i\Gamma_{j,J}^\dagger \Gamma_{k,K} &= i e^{i \pi \left( n_j + m_{j,J} \right)}e^{-i\pi \left(n_k + m_{k,K} \right)}.
\end{align}
There is no $\theta_\rho$ dependence in Eq.~\eqref{eq:neutral-product} because we have taken the limit ${K_\rho \to \infty}$.  We will return to this point at the end of Section~\ref{sec:comparison}.

We note several important features of Eq.~\eqref{eq:neutral-product}, all of which are discussed in more detail in Appendix~\ref{app:zero-mode}.  (1) The operators $n_{j/k}$, $m_{j/k,J/K}$ are integer-valued, thus the eigenvalues of $i\Gamma_{j,J}^\dagger\Gamma_{k,K}$ are $\pm 1$.
(2) $i\Gamma_{j,J}^\dagger\Gamma_{k,K}$ acts on the topologically protected parity eigenstates $\ket{\pm}$ of Eq.~\eqref{eq:parity-eigenstates} exactly as expected for bilinears of the Majorana operators $\gamma$ reviewed in the introduction.  (3) $i\Gamma_{j,J}^\dagger \Gamma_{k,K}$ commutes with all local operators.  Points (1-3) imply that in the limit $K_\rho\to \infty$, $i\Gamma_{j,J}^\dagger\Gamma_{k,K}$ can be identified with the MZM parity.  To emphasize this point, throughout the remainder of the paper we will write
\begin{align}\label{eq:neutral-product1}
    i\Gamma_{j,J}^\dagger\Gamma_{k,K} &= i\gamma_{j,J}\gamma_{k,K},
\end{align}
where $i\gamma_{j,L}\gamma_{j,R} \ket{\pm}_j = \pm \ket{\pm}_j.$  The correlation function Eq.~\eqref{eq:corr-function} thus reduces to
\begin{align}\label{eq:neutral-product1}
    \langle T_\tau \Gamma_{j,J}^\dagger(\tau_1)\Gamma_{k,K}(\tau_2)\rangle &= e^{-E_C|\tau_1-\tau_2|} \gamma_{j,J}\gamma_{k,K}.
\end{align}

Equations~\eqref{eq:neutral-product}-\eqref{eq:neutral-product1} establish a correspondence between MZM parity operators in number-conserving and mean-field approaches (see also Section~\ref{sec:comparison}). While fermion operators couple to both $\phi$ and $\theta_-$ degrees of freedom, the parity operator $i\Gamma_{j,J}^\dagger\Gamma_{k,K}$ is neutral and commutes with all local operators. Thus, degenerate ground states of the system (encoded in terms of MZM parity operators) cannot be distinguished by any local operator.

%%%
\section{Tunneling Measurement}\label{sec:mst}

We now review the tunneling measurement of MZM parity.  The basic idea is depicted in Fig.~\ref{fig:setup}.  Two bare semiconductor regions are separated by tunnel barriers from an intermediate quantum dot, {\it e.g.}, between $x_{2,R}$ and $x_{3,L}$. The measurement protocol involves lowering tunneling barriers and increasing the amplitude for virtual tunneling of an electron between the quantum dot and Majorana island (we assume that the charging energy is large so that there is still a charge gap in the system suppressing real single-electron tunneling processes). The relevant charge fluctuation processes involve an electron tunneling in and out of the Majorana island either through the same MZM, or in through one and out through the other. As a result, one finds a MZM parity-dependent energy shift of the combined qubit-quantum dot system, which can be used to infer the parity of the participating MZM pair. For simplicity, we focus on a parity measurement of two adjacent MZMs; the measurement can be generalized to other MZM pairs with the use of coherent links (floating topological superconducting islands in a fixed parity state) or by modifying the geometry of the qubit, as discussed at length in Ref.~\onlinecite{Karzig17}.

Following the above outlined idea, we now derive the measurement-induced energy shift using our particle-number-conserving formalism. The dot-Majorana island tunneling Hamiltonian can be written as
\begin{align}
H_t &= \sqrt{\ell}c_d^\dagger \left(t_{j,J} \psi(x_{j,J}+J\ell) +  t_{k,K}  \psi(x_{k,K}+K\ell)\right) + h.c.
\end{align}
where $c_d$ is the annihilation operator for the quantum dot and $t_{j,J}$ is the tunneling amplitude for an electron to tunnel into the semiconductor at $\psi(x_{j,J}+J\ell).$
The semiconductor electrons at the boundaries can be expanded as ${\psi(x_{j,J}+J\ell)= \Gamma_{j,J}/\sqrt{\ell}+ \dots}$, so that for sufficiently low temperatures (where the energy scale is set by the level spacing of the bare semiconductor region) $H_t$ becomes
\begin{align}\label{eq:Ht}
    H_t &=t_{j,J}c_d^\dagger \Gamma_{j,J} + t_{k,K} c_d^\dagger \Gamma_{k,K} + h.c.
\end{align}
Note that unlike the previous works~\onlinecite{Fu10,vanHeck16,Karzig17}, Eq.~\eqref{eq:Ht} uses the number-conserving expression for the fermionic zero mode $\Gamma_{k,K}$, rather than writing $H_t$ in terms of Majorana operators $\gamma_{k,K}$.

Odd orders in $H_t$ necessarily change the charge of the island and thus are exponentially suppressed by a large charge gap $~E_C \gg T$ (for $N_g=0$). Using imaginary-time path-integral formalism, one can derive the second order tunneling action to find
\begin{widetext}
\begin{align}
    S_t^{(2)} &= \frac{1}{2} \int_0^\beta d\tau_1 d\tau_2 \left(t_{j,J} c_d^\dagger(\tau_1) \Gamma_{j,J}(\tau_1) + t_{k,K} c_d^\dagger(\tau_1) \Gamma_{k,K}(\tau_1) + h.c. \right)  \left( t_{j,J} c_d^\dagger(\tau_2) \Gamma_{j,J}(\tau_2) + t_{k,K} c_d^\dagger (\tau_2)\Gamma_{k,K}(\tau_2) + h.c.\right) .
\end{align}
Averaging over the charging energy, we have
\begin{align}
 \left\langle S_t^{(2)}\right\rangle_C= \frac{1}{2} \int_0^\beta  d\tau_1 & \int_0^{\beta} d\tau_2  \Big\{ \Big( |t_{j,J}|^2 \langle T_\tau \Gamma_{j,J}(\tau_1)\Gamma_{j,J}^\dagger(\tau_2) \rangle_C + |t_{k,K}|^2 \langle T_\tau \Gamma_{k,K}(\tau_1)\Gamma_{k,K}^\dagger(\tau_2)\rangle_C \notag
\\ &+ t_{j,J} t_{k,K}^* \langle T_\tau \Gamma_{j,J}(\tau_1)\Gamma_{k,K}^\dagger(\tau_2)\rangle_C + t_{j,J}^* t_{k,K} \langle T_\tau \Gamma_{k,K}(\tau_1)\Gamma_{j,J}^\dagger(\tau_2)\rangle_C \Big) c_d^\dagger(\tau_1)c_d(\tau_2)
+ (\tau_1\leftrightarrow \tau_2) \Big\}
\\ =- \frac{1}{2} \int_0^\beta d\tau_1 d\tau_2 \Big\{ ~&e^{-E_C |\tau_1-\tau_2|}\Big( |t_{j,J}|^2 + |t_{k,K}|^2+2 \text{Im}[t_{j,J}^* t_{k,K}] i\gamma_{j,J}\gamma_{k,K}\Big) c_d^\dagger(\tau_1)c_d(\tau_2)+ (\tau_1\leftrightarrow \tau_2)
\Big\},
\end{align}
where in the last equality we have used Eq.~\eqref{eq:corr-function}.  In the limit $T\ll E_C$, we can take $\beta=1/T \to \infty$ so that
\begin{align}\label{eq:action-2}
    \left\langle S_t^{(2)}\right\rangle_C &= -2 \frac{|t_{j,J}|^2 + |t_{k,K}|^2+2 \text{Im}[t_{j,J}^* t_{k,K}] i\gamma_{j,J}\gamma_{k,K}}{E_C} \int_0^\beta  d\tau\, c_d^\dagger (\tau)c_d (\tau) +\mathcal{O}\left(E_C^{-2}\right),
\end{align}
\end{widetext}
see Appendix~\ref{app:mst} for details.  The effective tunneling Hamiltonian is thus
\begin{align} \label{eq:Heff}
    H_\text{eff} &= - 2\frac{|t_{j,J}|^2 + |t_{k,K}|^2+2 \text{Im}[t_{j,J}^* t_{k,K}] i\gamma_{j,J}\gamma_{k,K}}{E_C} c_d^\dagger c_d .
\end{align}
Higher orders in perturbation theory modify the parity-dependent energy splitting, but do not change the structure of Eq.~\eqref{eq:Heff}.  Thus, our number-conserving formalism has recovered the essential result from Ref.~\onlinecite{Karzig17} that tunneling results in a parity-dependent energy shift of the joint state of the quantum dot and the qubit.
The MZM parity can then be readout by probing the quantum dot ground state, {\it e.g.}, through spectroscopy, charge sensing, or differential capacitance~\cite{Karzig17}.

It is worth noting that noisy measurement or insufficient integration time could result in a partial projection of the MZM parity state.  Errors in the braiding phase implemented with a measurement-based protocol, reviewed below, will be bounded from below by measurement errors.  Therefore, topological protection is only achievable provided measurement errors are sufficiently suppressed.  Measurement errors warrant further consideration, but are independent of number-conserving effects and are thus beyond the scope of the current analysis.

%%%
\section{Implications for braiding}\label{sec:braiding}

The motivating question for this paper is whether number conservation in a topological superconductor introduces non-universal corrections to the MZM braiding phase.  We argue this is not the case in the context of measurement-based braiding.

Measurement-based braiding replaces physically moving MZMs with a sequence of projective parity measurements~\cite{Bonderson08b, Bonderson08c}.  This protocol utilizes the ancilla Hilbert space provided by encoding a qubit in six, rather than four, MZMs.  Mathematically, a measurement projects the MZM pair $i\gamma_j\gamma_k$ into a definite parity state.  The even and odd parity projectors are given by
\begin{align}
\Pi_\pm^{(jk)} &= \frac{1\pm i\gamma_j\gamma_k}{2}.
\end{align}
Recall that braiding MZMs $j$ and $k$ corresponds to the operator $R^{(jk)}$ given in Eq.~\eqref{eq:braid}. Let us encode the qubit state in MZMs $h$, $i$, $j$, and $k$, while $a$ and $b$ correspond to the ancilla MZM pair.  Then, $R^{(jk)}$ can be related to a sequence of even parity projections:
\begin{align}
\Pi_+^{(ab)} \Pi_+^{(aj)} \Pi_+^{(ak)} \Pi_+^{(ab)} &\propto R^{(jk)}\Pi_+^{(ab)}.
\end{align}
The above follows straightforwardly from Eq.~\eqref{eq:Maj-com}.  Note that each projector changes which four MZMs encode the qubit state, but does not collapse the encoded information.

While it is not in general possible to guarantee the outcome of a measurement ({\it e.g.}, whether $\Pi_+$ or $\Pi_-$ is applied), this complication can be circumvented by employing ``forced measurement"~\cite{Bonderson08b}.  If the wrong measurement outcome is obtained, simply repeat the previous parity measurement in the sequence, then re-attempt the desired measurement.   This repeat-until-success protocol does not change the relative phase implemented by the sequence, and on average requires two repeated measurements.   Reference~\onlinecite{Zhang16} considered how forced measurement may be circumvented by appropriately modifying the software tracking the measurement outcomes, while Ref.~\onlinecite{Knapp16} investigated the tradeoff between forced measurement and adiabatically tuning MZM couplings.  Reference~\onlinecite{Tran19} further investigated how to minimize the number of necessary MZM measurements for Clifford gates.

 The previous section demonstrated that number conservation only affects the tunneling measurement of MZM parity at energies on the order of $\mathcal{O}\left(E_C\right)$, and thus at low temperatures $T\ll E_C$ results in exponentially suppressed corrections $\mathcal{O}\left(e^{-E_C/T}\right)$.  Therefore, the underlying arguments of measurement-based braiding are unaltered by the number-conserving analysis of this paper.
Essentially, measurement-based braiding relies on the ability to project a pair of MZMs to the desired parity eigenstate.  Errors in this protocol arise from residual hybridization of MZMs.  Generally, MZM hybridization is exponentially suppressed in the energy gap over the temperature, and in the distance separating the MZMs over the correlation length of the topological superconductor.  When this is the case, the resulting braiding phase errors in a measurement-only protocol are similarly small and the protocol is topologically protected.

%%%
\section{Comparison to previous results}\label{sec:comparison}

We now discuss and compare our results with the previous works on this subject~\cite{Fu10,Gangadharaiah11, Lobos12,vanHeck16,Karzig17,Kim,Snizhko18}.  
The mean field equivalent of our bosonized analysis is to suppress superconducting phase fluctuations by replacing the field $\sqrt{2}\theta_\rho$ with a scalar quantity $\Phi$.  In this case, the pairing Hamiltonian becomes
\begin{align}
H_P^\text{MF} &= \sum_j \frac{\Delta_P}{2\pi a} \int_{x_{j,L}}^{x_{j,R}} dx \cos(2\theta-\Phi),
\end{align}
and no longer commutes with the number operator $N$.   Equation~\eqref{eq:gs-projection} is modified to
\begin{align}
    \Gamma_{j,J}^\text{MF}\to e^{i \frac{\Phi}{2}}e^{-i \pi \left(n_j + m_{j,J} \right)},
\end{align}
where $n_j$, $m_{j,J}$ are both integer-valued operators.  When ${\Phi=0}$, $\Gamma_{j,J}^\text{MF}$ is Hermitian and commutes with all bulk operators, therefore it can be identified with the Majorana operator $\gamma_{j,J}$ as established in Refs.~\onlinecite{Fidkowski11, Lobos12, Clarke13}.  %Reference~\onlinecite{Snizhko18} made this analysis number-conserving in the context of proximitized fractional quantum Hall edges by including a superconducting phase operator in the pairing Hamiltonian; they further considered the effect of a microscopically derived charging energy.

For Coulomb-blockaded Majorana islands, 
previous works~\cite{Fu10,vanHeck16,Karzig17} have used a phenomenological form of the Majorana tunneling Hamiltonian, 
%in the context of Coulomb-blockaded islands   
\begin{align}
\tilde{H}_t &= t c_d^\dagger \gamma e^{-i\hat{\Phi}/2} + h.c.
\end{align}
where $\hat{\Phi}$ is fluctuating superconducting phase that satisfies the commutation relation $[\hat{\Phi}, N]=2i$ with $N$ being the total charge of the island.  
By comparing with Eqs.~\eqref{eq:theta0} and \eqref{eq:Ht}, one may notice that $e^{i\hat{\Phi}/2}$ is similar to the dependence on $e^{i\theta_\rho/\sqrt{2}}$ in $\Gamma$. However, $\theta_\rho$ is dual to $N_\text{sc}$ rather than $N=N_\text{sc}+N_\text{sm}$, i.e. this operator adds a charge to the superconductor in contrast to a total charge between the superconductor and semiconductor. Thus, Majorana tunneling processes in general act on both topological and non-topological degrees of freedom. However, as we show above parities $\Gamma_{j,J}^\dagger \Gamma_{k,K}$ couple only to topological degrees of freedom (up to exponentially small corrections $O(e^{-E_C/T})$).  

%Reference~\onlinecite{Snizhko18} used a  number-conserving analysis in the context of proximitized fractional quantum Hall edges.

The differences between $\Gamma_{j,J}$ and $\Gamma_{j,J}^\text{MF}$ connect naturally to the concerns raised by Refs.~\onlinecite{Lin17,Lin18}.  In their case, the number-conserving version of the Majorana operator included a Cooper pair in its definition, and thus seems reminiscent of the dependence on $e^{i\theta_\rho(x_{j,J})/\sqrt{2}}$ in $\Gamma_{j,J}$.  However, their concern that the Cooper pair would introduce non-universal corrections to the braiding phase does not occur in our scenario.  Indeed, by neglecting spatial fluctuations in $\theta_\rho$ (and taking the limit $K_\rho\to \infty$), one can show that the $\theta_\rho$ dependence drops out of the neutral product $\Gamma_{j,J}^\dagger \Gamma_{k,K}$.  Temporal fluctuations in $\theta_\rho$ do not modify the tunneling measurement, as the charging energy effectively sets the times equal in $S_t^{(2)}$, so that the measurement only couples to the MZM parity.  Thus, for temperatures $T\ll E_C$, the tunneling-based parity measurement is not affected by imposing number conservation.

One might worry that our conclusions would change if we keep $K_\rho$ finite so that there are spatial fluctuations in $\theta_\rho.$  In Appendix~\ref{app:fluctuations}, we argue that for $K_\rho$ finite, the correlation function in Eq.~\eqref{eq:corr-function} becomes
\begin{align}
    &\langle T_\tau \Gamma^\dagger_{k,K}(\tau_1)\Gamma_{j,J}(\tau_2)\rangle \notag
    \\ &= e^{-E_C|\tau_1-\tau_2|}e^{ -\frac{1}{4}\langle\left[\theta_\rho(x_{j,J})-\theta_\rho(x_{k,K}) \right]^2 \rangle} \gamma_{j,J}\gamma_{k,K},
\end{align}
which in turn modifies the effective tunneling Hamiltonian to be
\begin{align} \label{eq:modified}
    H_\text{eff}& = - \frac{|t_{j,J}|^2 + |t_{k,K}|^2}{E_C} c_d^\dagger c_d\notag
    \\  +&e^{ -\frac{1}{4}\langle\left[\theta_\rho(x_{j,J})-\theta_\rho(x_{k,K}) \right]^2 \rangle} \frac{2\text{Im}[t_{j,J}^* t_{k,K}] i\gamma_{j,J}\gamma_{k,K}}{E_C} c_d^\dagger c_d .
\end{align}
The factor $e^{ -\frac{1}{4}\langle\left[\theta_\rho(x_{j,J})-\theta_\rho(x_{k,K}) \right]^2 \rangle}\leq 1$ saturates the bound when $K_\rho\to \infty,$ and otherwise reduces the measurement visibility (decays algebraically) when $K_\rho$ remains finite (the exact $K_\rho$ dependence is sensitive to which measurement is being performed), see Eq.~\eqref{eq:modified-connect}.   Thus, our results indicate that spatial quantum phase fluctuations in the superconductor reduce the measurement visibility, in addition to affecting the degeneracy splitting of the qubit states as reported earlier in Ref.~\onlinecite{Fidkowski11}.  This reduction in the measurement visibility may be particularly important for two-qubit measurements, for which the gap separating the ground state and first excited state in a fixed parity sector is reduced from $\mathcal{O}\left(E_C\right)$ for a single-qubit measurement to $\mathcal{O}\left(t^2/E_C\right)$.~\cite{Karzig17}  For the measurements proposed in Ref.~\onlinecite{Karzig17}, reduced visibility requires a longer integration time to achieve the same measurement accuracy, and can become problematic if the integration time becomes comparable to the qubit coherence times.  

%%%
\section{Conclusions}\label{sec:conclusions}

In this paper, we employed a number-conserving bosonized formalism to study 1D topological superconductors formed from semiconductor-superconductor heterostructures.  We demonstrated the presence of fermionic zero modes localized to the ends of a proximitized nanowire, and related these zero modes to the MZM parity operator.  We carefully considered the effect of tunnel coupling between the proximitized nanowire and an adjacent quantum dot, and showed that the combined system exhibits a parity-dependent energy shift independent of the topological state of the rest of the qubit, up to exponentially suppressed corrections from higher energy processes.  Finally, we showed that number-conserving corrections do not affect projective parity measurements and, as a result, measurement-based braiding operations are topologically protected.

Our findings contrast the conjecture by Refs.~\onlinecite{Lin17,Lin18} that number conservation could introduce non-universal corrections to the MZM braiding phase in a topological superconductor.  The critical step in our argument is that while the form of the fermionic zero mode $\Gamma_{j,L/R}$ localized to the left/right end of proximitized wire $j$ is modified in our number-conserving formalism as compared to a mean-field analysis, the relevant quantity $i\Gamma^\dagger_{j,J}\Gamma_{k,K}$ can still be identified with the mean-field MZM parity. Thus we affirm the potential of Majorana-based qubits to achieve topologically protected Clifford gates through braiding.

Previous studies have investigated the effect of quantum fluctuations in the superconductor on the MZM hybridization energy~\cite{Fidkowski11}.  Here, we have extended this analysis to the tunneling-based MZM parity measurement and have shown that spatial fluctuations can reduce the measurement visibility, in addition to the previously identified effects.

 Understanding how different noise sources affect MZM parity measurements is an interesting open question. The bosonized particle-number formalism utilized here provides a well-developed framework for investigating these effects.  Perturbation theory, for instance in gate voltage fluctuations coupling to density, can be straightforwardly applied to understand how this noise further reduces measurement visibility.  Additionally, the analysis could be extended to translate reduced visibility into fidelity estimates for measurement-based braiding by specifying the readout method ({\it e.g.}, charge sensing or differential capacitance).  As experimental progress in tuning semiconductor-superconductor nanowires into the topological phase continues to improve~\cite{Lutchyn_review}, such questions become of increasing practical importance.

%%%
\section*{Acknowledgments}

We are grateful to Torsten Karzig, Chetan Nayak, and Dmitry Pikulin for stimulating discussions. C.K. acknowledges support from the NSF GRFP under Grant No. DGE $114085$ and from the Walter Burke Institute for Theoretical Physics at Caltech.  Part of this work was performed at the Aspen Center for Physics, which is supported by National Science Foundation grant PHY-1607611.  

%%%
\appendix

%%%
\section{Zero-mode solutions}\label{app:zero-mode}

The general normal mode expansions for a Luttinger liquid are~\cite{Giamarchi04}
\begin{align}
\phi(x) &= \phi^0  - \frac{i\pi\sqrt{K}}{\ell}  \sum_{p\neq 0} \sqrt{ \frac{\ell|p|}{2\pi} } \frac{e^{-ipx-a|p|/2}}{p} \left( b_p^\dagger + b_{-p} \right)
\\ \theta(x) &= \theta^0 + \frac{i\pi}{\ell\sqrt{K}} \sum_{p\neq 0}  \sqrt{ \frac{\ell |p|}{2\pi} } \frac{e^{-ipx-a|p|/2}}{|p|} \left( b_p^\dagger - b_{-p} \right) \,,
\end{align}
where $a$  is the short-distance cutoff.
For the bare semiconductor segment residing at the $J$th side of the $j$ proximitized wire, we write the fields as $\phi_{j,J}$ and $\theta_{j,J}$, and impose boundary conditions $\theta_{j,J}(0)=\theta^0_{j,J}$ and $\phi_{j,J}(J\ell) = \phi_{j,J}^0.$  This sets $b_p=-b_{-p}$ and $p=\pi \left(k+\frac{1}{2}\right)/\ell$ in the above expansions, resulting in Eqs.~(\ref{eq:phi}-\ref{eq:theta}).  Their commutator is given by
\begin{widetext}
\begin{align}
    [\phi_{j,J}(x),\theta_{j,J}(y)] &= [\phi^0_{j,J},\theta^0_{j,J}]-i 4 \sum_{k=0}^\infty \frac{\cos\left([2k+1]\frac{\pi x}{2\ell}\right)\sin\left([2k+1]\frac{\pi y}{2\ell}\right)}{2k+1}
    \\ &= [\phi^0_{j,J},\theta^0_{j,J}]-i 2 \sum_{k=0}^\infty \frac{\sin\left([2k+1]\frac{\pi(x+y)}{2\ell}\right)-\sin\left([2k+1]\frac{\pi(x-y)}{2\ell}\right)}{2k+1}
    \\ &= [\phi^0_{j,J},\theta^0_{j,J}]-i \frac{\pi}{2} \left(\text{Sign}(x+y)-\text{Sign}(x-y)\right), \label{eq:A5}
\end{align}
where Eq.~\eqref{eq:A5} follows from the identity
\begin{align}
    \sum_{k=0}^\infty \frac{\sin\left([2k+1]x\right)}{2k+1} &= \frac{\pi}{4}\text{Sign}(x).
\end{align}
When $J=L$, $x,y<0$ and
\begin{align}
    -i\frac{\pi}{2}\left(\text{Sign}(x+y)-\text{Sign}(x-y)\right) &= i \frac{\pi}{2} \left( 1 + \text{Sign}(x-y)\right) = i \pi \Theta(x-y),
\end{align}
which implies $[\phi_{j,L}^0,\theta^0_{j,L}]=0$.  When $J=R$, $x,y>0$ and
\begin{align}
  -i\frac{\pi}{2}\left(\text{Sign}(x+y)-\text{Sign}(x-y)\right) &=-i\frac{\pi}{2}\left( 1-\text{Sign}(x-y)\right)
= -i \pi \Theta(y-x).
\end{align}
\end{widetext}
Therefore, the for the right segment, $[\phi^0_{j,R},\theta^0_{j,R}]= i \pi $, in order to satisfy Eq.~\eqref{eq:commutator}.  Note that the bosonic operators $b_k$ for different bare semiconductor segments commute, thus Eq.~\eqref{eq:commutator} implies that the zero modes more generally satisfy
\begin{align}
    [\phi_{j,J}^0,\theta_{k,K}^0]&= i \pi \Theta(j-k+J/2).
\end{align}

\subsection{Derivation of $\Gamma_{j,J}$ }\label{app-sec:commutator}

In this section, we derive a charge-one fermionic zero mode of $H_\text{bare}$.  Our derivation closely follows that of Refs.~\onlinecite{Fidkowski12,Clarke13}.

From the normal mode expansions in Eqs.~(\ref{eq:phi}-\ref{eq:theta}) we split the fields $\phi_{j,J}$ and $\theta_{j,J}$ into zero-mode and higher harmonic pieces:
\begin{align}
    \phi_{j,J}(x) &= \phi_{j,J}^0 + \sum_{k=0}^\infty \phi^k(x)
    \\ \theta_{j,J}(x) &= \theta_{j,J}^0 + \sum_{k=0}^\infty \theta^k(x).
\end{align}
It is convenient to introduce fields $\varphi_{r/l}$  defined by
\begin{align}\label{eq:chiral-simple}
\varphi_{r/l}(x) &=\theta^0_{j,J} \mp \phi^0_{j,J}+ \sum_{k=0}^\infty \left( K \theta^k(x) \mp \phi^k(x)\right)
\\ &= \varphi_{r/l}^0 \mp \sum_{k=0}^\infty \varphi_{r/l}^k(x) \label{eq:chiral-simple1}
\end{align}
which satisfy
\begin{align}\label{eq:A13}
    [H_\text{bare},\varphi_{r/l}(x)]&= \mp i v \partial_x \varphi_{r/l}(x).
\end{align}
The Heisenberg equation therefore implies that $\varphi_{r/l}$ are chiral:
\begin{align}
    \partial_t \varphi_{r/l} &= i [H_\text{bare},\varphi_{r/l}]= \pm v \partial_x \varphi_{r/l}.
\end{align}
When $K=1$ the right/left-moving electrons can be written in terms of $\varphi_{r/l}$ as
\begin{align}
    \psi_{r/l}(x)&\sim e^{i\theta(x)\mp \phi(x)} =\lim_{K\to 1} e^{i\varphi_{r/l}(x)}.
\end{align}
(When $K\neq 1$, $e^{i\varphi_{r/l}}$ mixes $\psi_r$ and $\psi_l$.)

We can construct a zero mode of the full many-body spectrum by considering superpositions of $e^{\pm i\varphi_{r/l}}$.  In particular, Eq.~\eqref{eq:A13} implies
\begin{align}
    [H_\text{bare},\int_0^{J\ell}dx e^{i\varphi_{r/l}(x)}]&=\mp iv \left(e^{i\varphi_{r/l}(J\ell)}-e^{i\varphi_{r/l}(0)}\right)
    \\ &= \mp iv \left( e^{i \theta^0_{j,J}} - e^{\mp i \phi^0_{j,J}}\right),
\end{align}
where in the last line we have used the boundary conditions $\phi^k(J\ell)=0$ and $\theta^k(0)=0.$  Similarly,
\begin{align}
    [H_\text{bare},\int_0^{J\ell}dx e^{-i\varphi_{r/l}(x)}]&=
    \mp iv \left( e^{-i \theta^0_{j,J}} - e^{\pm i \phi^0_{j,J}}\right).
\end{align}
Therefore, we have that the superposition
\begin{widetext}
\begin{align}\label{eq:Gammadef}
    \Gamma_{j,J} &= \frac{J}{\ell} \int_0^{J\ell} dx \left\{ e^{i\varphi_r}+ e^{-i2\phi_{j,J}^0}e^{i\varphi_l} + e^{i2\theta_{j,J}^0-i2\phi_{j,J}^0}e^{-i\varphi_r} + e^{i2\theta_{j,J}^0}e^{-i\varphi_l} \right\}
\end{align}
\end{widetext}
is a zero mode of $H_\text{bare}$:
\begin{align}
    [H_\text{bare},\Gamma_{j,J}]&=0.
\end{align}

When $K=1$, the dependence on $\theta_{j,J}^0$ and $\phi_{j,J}^0$ can be written as ${e^{i2\theta^0_{j,J}} = \psi_r(0)\psi_l(0)}$ (Andreev boundary conditions) and ${e^{-i2\phi^0_{j,J}}=\psi_l^\dagger(J\ell)\psi_r(J\ell)}$ (normal boundary conditions).  In this case $\Gamma_{j,J}$ can be expressed in terms of left and right moving electrons as
\begin{widetext}
\begin{align}\label{eq:GammaK1}
   \lim_{K\to 1} \Gamma_{j,J} &= \frac{J}{\ell}\int_0^{J\ell}dx\left\{ \psi_r(x)+ \left( \psi_l^\dagger(J\ell)\psi_r(J\ell)\right) \psi_l(x) +\left(\psi_r(0)\psi_l(0)\right) \left(\psi_l^\dagger(J\ell)\psi_r(J\ell)\right) \psi_r^\dagger(x) + \left(\psi_r^{}(0)\psi_l(0)\right)\,\psi_l^\dagger(x) \right\}.
\end{align}
\end{widetext}
Equation~\eqref{eq:GammaK1} makes it especially apparent that $\Gamma_{j,J}$ is both charge-one and fermionic.  This also holds in the case ${K\neq 1}$, which becomes more obvious after taking the ground state projection.

To project $\Gamma_{j,J}$ to the ground state subspace, we first rewrite Eq.~\eqref{eq:Gammadef} using Eq.~\eqref{eq:chiral-simple1}:
\begin{align}\label{eq:A23}
\Gamma_{j,J}  &=  e^{i\theta_{j,J}^0 - i\phi_{j,J}^0} \int_0^{J\ell}\frac{J dx}{\ell} \left\{ e^{i\sum_k \varphi_r^k} + e^{i\sum_k \varphi_l^k} + h.c. \right\}.
\end{align}
The integrand only depends on the operators $b_k$, $b_k^\dagger$- all zero-mode dependence has been pulled in front.  Therefore, after projecting to the ground state subspace, the integrand contributes an unimportant constant~\cite{Clarke13} and we arrive at the expression used throughout the main text
\begin{align}
\Gamma_{j,J}&= e^{i\theta_{j,J}^0-i\phi_{j,J}^0}.
\end{align}

\subsection{Fermionic anticommutation}\label{app-sec:anticommutation}

Fermionic anticommutation of the zero modes $\Gamma_{j,J}$ follows straightforwardly from Eq.~\eqref{eq:zm-commutator}, $\Gamma_{j,J}^\dagger \Gamma_{j,J}=1$, and ${e^A e^B = e^B e^A  e^{[A,B]}}$ when $[A,[A,B]]=[B,[A,B]]=0$:
\begin{widetext}
\begin{align}
    \Gamma_{j,J}^\dagger \Gamma_{k,K} &= e^{-i\theta_{j,J}^0 +i\phi_{j,J}^0}e^{i\theta_{k,K}^0 -i\phi_{k,K}^0}
 %   \\ &= e^{-i\theta_{j,J}^0} e^{i\phi_{j,J}^0} e^{i\theta_{k,K}^0} e^{-i\phi_{k,K}^0} e^{-\frac{1}{2}[\theta_{j,J}^0,\phi_{j,J}^0]-\frac{1}{2}[\theta_{k,K}^0,\phi_{k,K}^0]}
 %   \\ &=e^{-i\theta_{j,J}^0} e^{i\theta_{k,K}^0} e^{i\phi_{j,J}^0} e^{-i\phi_{k,K}^0} e^{-[\phi_{j,J}^0,\theta_{k,K}^0]} e^{-\frac{1}{2}[\theta_{j,J}^0,\phi_{j,J}^0]-\frac{1}{2}[\theta_{k,K}^0,\phi_{k,K}^0]}
 %   \\ &= e^{i\theta_{k,K}^0} e^{-i\phi_{k,K}^0}e^{-i\theta_{j,J}^0}  e^{i\phi_{j,J}^0} e^{-[\theta_{j,J}^0,\phi_{k,K}^0]}e^{-[\phi_{j,J}^0,\theta_{k,K}^0]} e^{-\frac{1}{2}[\theta_{j,J}^0,\phi_{j,J}^0]-\frac{1}{2}[\theta_{k,K}^0,\phi_{k,K}^0]}
\\&=e^{i\theta_{k,K}^0 -i\phi_{k,K}^0}e^{-i\theta_{j,J}^0 +i\phi_{j,J}^0}e^{-[\theta_{j,J}^0,\phi_{k,K}^0]}e^{-[\phi_{j,J}^0,\theta_{k,K}^0]}
    \\ &= \Gamma_{k,K}\Gamma_{j,J}^\dagger e^{i\pi \Theta\left(k-j+\frac{K}{2}\right)-i\pi \Theta\left(j-k+\frac{J}{2}\right)}
    \\ &=- \Gamma_{k,K}\Gamma_{j,J}^\dagger\left(1-\delta_{j,k}\delta_{J,K}\right) + \delta_{j,k}\delta_{J,K}\Gamma_{k,K}\Gamma_{j,J}^\dagger
\end{align}
\end{widetext}
It follows from here that $\{\Gamma_{j,J}^\dagger,\Gamma_{k,K}\}= 2\delta_{j,k}\delta_{J,K}$.
To see the anticommutation before ground state projection, use Eq.~\eqref{eq:A23} and note that the operators $e^{\pm i \varphi_{r/l}^k(x)}$ anticommute.

\subsection{Action on relative fermion parity eigenstates}\label{app-sec:relative-parity}

Next, we show that $i\Gamma_{j,J}^\dagger\Gamma_{k,K}$ acts on the relative fermion parity eigenstates of Eq.~\eqref{eq:parity-eigenstates} exactly as expected for Majorana bilinears.  For the same wire, when $K_\rho \to \infty$, ${\theta_{j,L}^0=\theta_{j,R}^0=\theta_j^0}$
\begin{align}
    i\Gamma_{j,L}^\dagger \Gamma_{j,R} &= i e^{-i\theta_{j}^0 +i\phi_{j,L}^0}e^{i\theta_{j}^0-i\phi_{j,R}^0}
    \\ &= i e^{i\left(\phi_{j,L}^0 - \phi_{j,R}^0\right)} e^{\frac{1}{2}\left([\phi_{j,L}^0,\theta_{j}^0]-[\phi_{j,R}^0,\theta_j^0] \right)}
    \\ &= e^{i \left(\phi_{j,L}^0-\phi_{j,R}^0 \right)}.
\end{align}
The fermion parity eigenstates for wire $k$ are
\begin{align}
    \theta_-^k &= \frac{\theta_\rho}{\sqrt{2}}-\theta_k^0
    \\ \ket{\pm}_k &= \frac{1}{\sqrt{2}} \left( \ket{\theta_-^k =0} \pm \ket{\theta_-^k=\pi} \right).
\end{align}
Using
\begin{align}
    e^{i \phi_{0,J}^j}\ket{\theta_-^k}&=\ket{\theta_-^k + \pi \Theta\left(j-k+J/2\right)}
\end{align}
and $\theta_-^k+2\pi=\theta_-^k$,
we have
\begin{align}
    i\Gamma_{j,L}^\dagger \Gamma_{j,R}\ket{\pm}_j &= e^{i\left( \phi_{j,L}^0-\phi_{j,R}^0\right)} \frac{1}{\sqrt{2}}\left(\ket{\theta_-^j=0}\pm \ket{\theta_-^j=\pi} \right)
    \\ &= \frac{1}{\sqrt{2}}\left(\ket{\theta_-^j=\pi}\pm \ket{\theta_-^j=0} \right)
    \\ &= \pm \ket{\pm}_j.
\end{align}

The neutral product of fermionic zero modes for different wires similarly act as Majorana bilinears.  Consider first ${J=K=M}$ and $j<k$:
\begin{align}
    i\Gamma_{j,M}^\dagger\Gamma_{k,M}&=i e^{i\left(\theta_k^0-\theta_j^0\right)} e^{i \left(\phi_{j,M}^0 - \phi_{k,M}^0\right)} e^{[\phi_{j,M}^0,\theta_k^0]}
    \\ &= i e^{i\left(\theta_k^0-\theta_j^0\right)} e^{i \left(\phi_{j,M}^0 - \phi_{k,M}^0\right)}
\end{align}
Note that
\begin{align}
    &i\Gamma_{j,L}^\dagger \Gamma_{k,L} \ket{\pm}_j \ket{\pm }_k
    \\ &=i\Gamma_{j,L}^\dagger \Gamma_{k,L} \frac{1}{2} \left(\ket{0}_j\ket{0}_k +\ket{\pi}_j\ket{\pi}_k \pm \ket{0}_j\ket{\pi}_k \pm \ket{\pi}_j\ket{0}_k \right)
    \\ &= i e^{i\left(\theta_0^k-\theta_0^j\right)}\frac{1}{2}\left(\ket{\pi}_j\ket{0}_k +\ket{0}_j\ket{\pi}_k \pm \ket{\pi}_j\ket{\pi}_k \pm \ket{0}_j\ket{0}_k \right)
    \\ &= \frac{i}{2} \left(-\ket{\pi}_j\ket{0}_k -\ket{0}_j\ket{\pi}_k \pm \ket{\pi}_j\ket{\pi}_k \pm \ket{0}_j\ket{0}_k  \right)
    \\ &= \pm i  \ket{\mp}_j\ket{\mp}_k.
\end{align}
Therefore, we have
\begin{align}
    &i\Gamma_{j,L}^\dagger\Gamma_{k,L} \frac{1}{\sqrt{2}} \left(\ket{+}_j\ket{+}_k \pm i\ket{-}_j\ket{-}_k \right)
    \\ & \frac{1}{\sqrt{2}} \left(i\ket{-}_j\ket{-}_k \pm\ket{+}_j\ket{+}_k \right)
    \\ & \pm \frac{1}{\sqrt{2}}\left(\ket{+}_j\ket{+}_k \pm i\ket{-}_j\ket{-}_k \right) .
\end{align}
The argument for $i\Gamma_{j,R}^\dagger\Gamma_{k,R}$ follows similarly, except $\theta_-^k$, rather than $\theta_-^j$, advances by $\pi$.

For opposite ends of different wires $(j<k)$ we have
\begin{align}
    i\Gamma_{j,L}^\dagger \Gamma_{k,R} &= - e^{i\left(\theta_k^0-\theta_j^0\right)} e^{i \left(\phi_{j,L}^0 - \phi_{k,R}^0\right)} e^{[\phi_{j,L}^0,\theta_k^0]}
    \\ &= -e^{i\left(\theta_k^0-\theta_j^0\right)} e^{i \left(\phi_{j,L}^0 - \phi_{k,R}^0\right)},
\end{align}
\begin{align}
    &i\Gamma_{j,L}^\dagger\Gamma_{k,R} \ket{\pm}_j\ket{\pm}_k\notag
    \\ &= i\Gamma_{j,L}^\dagger\Gamma_{k,R} \frac{1}{2} \left( \ket{0}_j\ket{0}_k + \ket{\pi}_j\ket{\pi}_k \pm \ket{0}_j\ket{\pi}_k \pm \ket{\pi}_j\ket{0}_k \right)
    \\ &= e^{i\left(\theta_0^k-\theta_0^j\right)} \frac{1}{2} \left( \ket{\pi}_j\ket{\pi}_k + \ket{0}_j\ket{0}_k \pm \ket{\pi}_j\ket{0}_k \pm \ket{0}_j\ket{\pi}_k \right)
    \\ &= \mp\ket{\mp}_j\ket{\mp}_k.
\end{align}
It follows that
\begin{align}
  & i\Gamma_{j,L}^\dagger\Gamma_{k,R} \frac{1}{\sqrt{2}}\left(\ket{+}_j\ket{+}_k \pm \ket{-}_j\ket{-}_k \right)
  \\ &= \pm  \frac{1}{\sqrt{2}}\left(\ket{+}_j\ket{+}_k \pm \ket{-}_j\ket{-}_k \right) .
\end{align}
Thus, we have shown all choices of $i\Gamma_{j,J}^{\dagger}\Gamma_{k,K}$ act on the relative fermion parity eigenstates exactly as expected for the MZM parity (for $K_\rho\to \infty$).

\subsection{Commutation with local operators}\label{app-sec:comm-local-op}

Topologically encoded information should be unobservable to any local operator.  We now demonstrate that $\Gamma_{j,J}$ commutes with all fermionic bilinears.  For simplicity, we focus on the ground state projected expression, Eq.~\eqref{eq:gs-projection}.

First, commutation with gradients and superconducting fields follows trivially. Thus, we want to show commutation with any term of the form $e^{i (a\theta(x)+b\phi(x))}e^{i(c\theta(x)+d\phi(x)}$, where $a,\,b,\,c\,$ and $d$ are $\pm 1$.  This reduces to demonstrating $[\Gamma_{j,J},e^{i2\phi(x)}]=[\Gamma_{j,J},e^{i2\theta(x)}]=0.$ These follow from
\begin{widetext}
\begin{align}
[e^{i\theta_{j,J}^0},e^{i2\phi(x)}]&=e^{i\theta_{j,J}^0+i2\phi(x)}\left( e^{\frac{1}{2}[2\phi(x),\theta_{j,J}^0]}- e^{\frac{1}{2}[\theta_{j,J}^0,2\phi(x)]}\right)=0
\\ [e^{i\phi_{j,J}^0},e^{i2\theta(x)}]&=e^{i\phi_{j,J}^0+i2\theta(x)}\left( e^{\frac{1}{2}[2\theta(x),\phi_{j,J}^0]}- e^{\frac{1}{2}[\phi_{j,J}^0,2\theta(x)]}\right) =0.
\end{align}
\end{widetext}
Therefore, $\Gamma_{j,J}$ commutes with all local operators.

%%%
\section{Correlation function derivation}\label{app:corr-function}

We now derive Eq.~\eqref{eq:corr-function}.  First, note that from the time-dependent expressions Eqs.~\eqref{eq:Gamma-time} and \eqref{eq:Gamma-time-dagger} we have two expressions for $\Gamma_{j,J}(\tau)$:
\begin{align}
    \Gamma_{j,J}(\tau) &= e^{-E_C\left(2N-2N_g +1\right)\tau}\Gamma_{j,J}(0)
    \\ &= \Gamma(0) e^{-E_C\left(2N-2N_g -1\right)\tau}.
\end{align}
The first expression is derived by solving the Heisenberg equation of motion of $\Gamma_{j,J}(\tau)$, while the second comes from solving the Heisenberg equation of motion for $\Gamma_{j,J}^\dagger(\tau)$ and taking the Hermitian conjugate.  Similarly,
\begin{align}
    \Gamma_{j,J}^\dagger(\tau) &= e^{E_C\left(2N-2N_g -1\right)\tau}\Gamma^\dagger(0)
    \\ &= \Gamma^\dagger(0)e^{E_C\left(2N-2N_g +1\right)\tau}.
\end{align}

Using these expressions, we have
\begin{widetext}
\begin{align}
     T_\tau \Gamma_{j,J}^\dagger(\tau_1)\Gamma_{k,K}(\tau_2)
     &=  \Theta(\tau_1-\tau_2) \Gamma_{j,J}^\dagger(\tau_1)\Gamma_{k,K}(\tau_2) - \Theta(\tau_2-\tau_1) \Gamma_{k,K}(\tau_2)\Gamma_{j,J}^\dagger(\tau_1)
     \\ &= \Theta(\tau_1-\tau_2) e^{E_C\left(2N-2N_g-1\right)\tau_1}\Gamma_{j,J}^\dagger(0) \Gamma_{k,K}(0) e^{-E_C\left(2N-2N_g-1\right)\tau_2} \notag
     \\ &- \Theta(\tau_2-\tau_2) e^{-E_C\left(2N-2N_g+1\right)\tau_2}\Gamma_{k,K}(0)\Gamma_{j,J}^\dagger(0)e^{E_C\left(2N-2N_g+1\right)\tau_1}
     \\ &= \gamma_{j,J}\gamma_{k,K} e^{E_C\left(2N-2N_g\right)\left(\tau_1-\tau_2\right)} e^{-E_C|\tau_1-\tau_2|}.
\end{align}
\end{widetext}
In the penultimate line, we used the fact that the $\Gamma_{k,K}(0)\Gamma_{j,J}^\dagger(0)$ is proportional to the MZM parity and therefore commutes with the number operator $N$.  Now, in the charging energy ground state,
\begin{align}
-\frac{1}{2} \leq \langle (N-N_g)\rangle_C \leq \frac{1}{2},
    %\langle  T_\tau \Gamma_{j,J}^\dagger(\tau_1)\Gamma_{k,K}(\tau_2)  \rangle_C &= \gamma_{j,J}\gamma_{k,K} \langle e^{E_C(2N-2N_g)(\tau_1-\tau_2)}\rangle_C e^{-E_C|\tau_1-\tau_2|}
\end{align}
therefore $\langle  T_\tau \Gamma_{j,J}^\dagger(\tau_1)\Gamma_{k,K}(\tau_2)  \rangle_C$ is always exponentially decaying in time.  We can simplify the problem by focusing on $N_g=0$, for which $\langle (N-N_g)\rangle_C =0$ and find Eq.~\eqref{eq:corr-function} of the main text
\begin{align}
    \langle  T_\tau \Gamma_{j,J}^\dagger(\tau_1)\Gamma_{k,K}(\tau_2)  \rangle_C &= \gamma_{j,J}\gamma_{k,K} e^{-E_C|\tau_1-\tau_2|}.
\end{align}

%%%
\section{Tunneling measurement details}\label{app:mst}

To arrive at Eq.~\eqref{eq:action-2}, we need to expand the product $c_d^\dagger(\tau_1)c_d(\tau_2)$ around $\tau_1=\tau_2$.  This can be achieved by switching variables to
\begin{align}
    S &= \frac{\tau_1+\tau_2}{2}\,,
& s &= \tau_1-\tau_2 \,,
\end{align}
so that
\begin{widetext}
\begin{align}
    &\int_0^\beta d\tau_1 \int_0^\beta d\tau_2\, e^{-E_C|\tau_1-\tau_2|} c_d^\dagger(\tau_1)c_d(\tau_2) = \int_0^\beta dS \int_{-2\min(S,\beta-S)}^{2\min(S,\beta-S)} ds \, e^{-E_C|s|} c_d^\dagger(S+\frac{1}{2}s)c_d(S-\frac{1}{2}s)
    \\ &= \int_{\tau_c}^{\beta-\tau_c} dS \int_{-2\min(S,\beta-S)}^{2\min(S,\beta-S)} ds \, e^{-E_C|s|} \left(c_d^\dagger(S)+ \frac{1}{2}s \partial_S c_d^\dagger(S) + \mathcal{O}(s^2) \right)\left(c_d(S)-\frac{1}{2} s \partial_S c_d(S) +\mathcal{O}(s^2)\right)
    \\ & \approx  \int_0^\beta dS \int_{-\infty }^{\infty} ds \, e^{-E_C|s|}  c_d^\dagger(S) c_d(S) + \frac{1}{2}\int_{-\infty}^{\infty} ds \, s e^{-E_C|s|} \int_0^\beta dS \left(\partial_S c_d^\dagger(S)\, c_d(S) + c_d^\dagger(S)\partial_S c_d(S) \right) \notag
    \\ &\quad +  \mathcal{O}\left(\int_{-\infty}^{\infty}ds\, s^2 e^{-E_C|s|}\right)
    \\ &= 2\frac{1}{E_C} \int_0^\beta dS c_d^\dagger(S)c_d(S) + \mathcal{O}\left( E_C^{-2}\right).
\end{align}
In the second line we assumed a short-time cutoff $\tau_c \sim \alpha$ in the  $S$-integral. We then assumed $E_C \tau_c \gg 1$ and extended the range of the $s$-integral in the third line.
Therefore,
\begin{align}
    \langle S_t^{(2)}\rangle_C &=- \left(|t_{j,J}|^2 + |t_{k,K}|^2 + 2 \text{Im} [t_{j,J}^* t_{k,K}] i\gamma_{j,J}\gamma_{k,K} \right) \int_0^\beta d\tau_1 \int_0^\beta d\tau_2 \,e^{-E_C|\tau_1-\tau_2|} c_d^\dagger(\tau_1)c_d(\tau_2) %+ \left(\tau_1\leftrightarrow\tau_2\right)
    \\ &=-2 \frac{|t_{j,J}|^2 + |t_{k,K}|^2 + 2 \text{Im} [t_{j,J}^* t_{k,K}] i\gamma_{j,J}\gamma_{k,K} }{E_C} \int_{0}^\beta d S c_d^\dagger(S)c_d(S) +\mathcal{O}\left( E_C^{-2}\right).
\end{align}
\end{widetext}
Importantly, each subsequent expansion in the difference ${\tau_1-\tau_2}$ contributes an additional factor of $E_C^{-1}$. Alternatively, the effective action could be derived by modeling the quantum dot as in Ref.~\onlinecite{Karzig17} to solve explicitly for the time dependence of the quantum dot operators $c_d$.  This contributes an additional term to the denominator of the quantum dot's charging energy.

%%%
\section{Effect of spatial fluctuations of $\theta_{\rho}$.}\label{app:fluctuations}

In this appendix, we discuss how our results effect of spatial fluctuations of $\theta_{\rho}$, {\it i.e.}, finite $K_\rho$. Let's consider first the case of a single wire and examine how the equal time correlator $\langle \Gamma_{j,K}^\dagger \Gamma_{j,J}\rangle$ changes when $\partial_x\theta_\rho\neq 0$:
\begin{align}\label{eq:new-corr}
    \left\langle \Gamma_{j,L}^\dagger \Gamma_{j,R}\right\rangle &= i \langle e^{\frac{i}{\sqrt{2}}\left(\theta_\rho(x_{j,R}) - \theta_\rho(x_{j,L})\right)} e ^{i\pi (m_{j,L}-m_{j,R} )}\rangle.
\end{align}
Just as it was useful to define a difference field $\theta_-^j$ for wire $j$, it is also useful to define an average field
\begin{align}
    \theta_+^j &=\frac{1}{2}\left( \frac{\theta_\rho}{\sqrt{2}} + \theta \right)
\end{align}
so that $\theta_\rho$ can be rewritten as
\begin{align}
    \frac{\theta_\rho}{\sqrt{2}} &= \theta_+ + \frac{\theta_-}{2}.
\end{align}
Given that $\theta_-$ is pinned by $\Delta_P$ to a spatially constant value for a given wire implies that the $\theta_-$ dependence drops out of Eq.~\eqref{eq:new-corr}:
\begin{align}
    \langle \Gamma_{j,L}^\dagger \Gamma_{j,R}\rangle &= i \left\langle e^{-i (\theta_+(x_{j,L})-\theta_+(x_{j,R}))} e^{i\pi (m_{j,L}-m_{j,R})}\right\rangle.
\end{align}
The $\theta_+$ fields (only defined in the proximitized wire section) decouple from the $m_{j,J}$ fields (defined in the bare semiconductor wire section), so the correlator can be factored.  As we have already shown that the term $e^{i\pi (m_{j,L}-m_{j,R})}=\gamma_{j,R}\gamma_{j,L}$, we have
\begin{align}
    \langle \Gamma_{j,L}^\dagger \Gamma_{j,R}\rangle &= i \left\langle e^{-i(\theta_+(x_{j,L})-\theta_+(x_{j,R}))}\right\rangle \gamma_{j,R}\gamma_{j,L}.
\end{align}

Finally, using the formula
\begin{align}
    \left\langle e^{i\left[\theta_+(x)-\theta_+(y)\right]}\right\rangle &= e^{-\frac{1}{2} \langle \left[\theta_+(x)-\theta_+(y)\right]^2\rangle },
\end{align}
we just need to evaluate $\langle (\theta_+(x)-\theta_+(y))^2\rangle.$  For a single wire, the action in terms of the $\theta_\pm$ fields is (neglecting the barrier and charging energy terms)
\begin{widetext}
\begin{align}
    S = \int d\tau \frac{1}{2\pi} \int dx &\Big\{ -2i \partial_\tau \theta_+ \partial_x \phi_+ -2i\partial_\tau \theta_- \partial_x \phi_- \notag
    \\ & + \left( 2v_\rho K_\rho + vK \right)\left( \left(\partial_x \theta_+\right)^2 + \frac{1}{4}\left(\partial_x \theta_-\right)^2\right) + \left(2 v_\rho K_\rho - vK\right)\left(\partial_x \theta_+\right)\left(\partial_x \theta_-\right) \notag
    \\ &+ \left(\frac{v_\rho}{2K_\rho} + \frac{v}{K} \right) \left(\frac{1}{4}\left(\partial_x \phi_+\right)^2 + \left(\partial_x \phi_-\right)^2 \right) + \left( \frac{v_\rho}{2K_\rho} - \frac{v}{K} \right) \left(\partial_x \phi_+\right)\left(\partial_x \phi_-\right) + \frac{\Delta_P}{\xi} \cos(2\theta_-) \Big\}.
\end{align}
\end{widetext}
where $\xi\sim v/\Delta_P$ is the coherence length which defines the short-range cutoff at the strong-coupling fixed point due to the pairing term. Note that the action is quadratic for the $\theta_+$ field.  If we take $\theta_-$ to be pinned from the cosine term, then we can neglect spatial and temporal fluctuations of $\theta_-$, so that the action decouples for the $\pm$ fields (temporal fluctuations contribute instanton terms, which result in an exponentially suppressed degeneracy splitting~\cite{Knapp17}).
Defining the coefficient of $(\partial_x\theta_+)^2$ as $K_+v_+$ and the coefficient of $(\partial_x\phi_+)^2$ as $v_+/K_+$, the resulting action for $\theta_+,\phi_+$ maps to a Luttinger liquid action, with effective Luttinger liquid parameter
\begin{align}
    K_+ &= 2\sqrt{2 K_\rho K}\sqrt{\frac{2v_\rho K_\rho + v K}{K v_\rho + 2K_\rho v}},
\end{align}
which in the limit of $K_\rho \gg 1$ becomes
\begin{align}
    K_+ \approx 2\sqrt{2 K_\rho K \frac{v_\rho}{v}}.
\end{align}

The correlator $\langle (\theta_+(x)-\theta_+(y))^2\rangle$ will therefore be given by that of a Luttinger liquid with Luttinger liquid parameter $K_+$~\cite{Giamarchi04}:
\begin{align}
    \langle[\theta_+(x_{j,R})-\theta_+(x_{j,L})]^2\rangle &= \frac{1}{2K_+}\log  \frac{(x_{j,R}-x_{j,L})^2 + \xi^2)}{\xi^2}.
\end{align}
Assuming the wire length is $L_\text{wire}$, this implies
\begin{align}
    \left\langle e^{i(\theta_+(x_{j,R})-\theta_+(x_{j,L}))}\right\rangle  &= \left( \frac{\xi^2}{L_\text{wire}^2+\xi^2}\right)^{\frac{1}{4 K_+}}
\end{align}
Reconnecting to Eq.~\eqref{eq:modified} in the main text, we have argued that for a single wire
\begin{align}\label{eq:modified-connect}
    e^{-\frac{1}{4} \langle \left[ \theta_\rho(x_{j,L})-\theta_\rho(x_{j,R})\right]^2\rangle} &=  e^{-\frac{1}{2}\langle \left[ \theta_+(x_{j,L})-\theta_+(x_{j,R})\right]^2\rangle}
    \\ &\approx \left( \frac{\xi}{L_\text{wire}} \right) ^{\sqrt{\frac{v}{32 K_\rho K v_\rho}}},
\end{align}
where in the last line we have taken the limit $L_\text{wire}\gg \xi$ and plugged in the definition of $K_+.$
As $K_\rho\to \infty$, the exponent approaches $0$ and the results are unaffected.  When $K_\rho$ remains finite, the above expression is smaller than $1$, and thus reduces the measurement visibility.

For multiple wires, the calculation changes somewhat, but the conclusion remains the same.  Assuming that for multiple wires the backbone contribution to the action is the dominant term, we can as a first approximation ignore the proximitized wires and find that the correlator is suppressed by a factor $ \left(\frac{\xi}{L_\text{wire}} \right)^{\frac{1}{4 K_\rho}}.$  The proximitized wires will add additional $K$ dependence.

%%%
%\bibliography{RLproject}

\begin{thebibliography}{58}%
\makeatletter
\providecommand \@ifxundefined [1]{%
 \@ifx{#1\undefined}
}%
\providecommand \@ifnum [1]{%
 \ifnum #1\expandafter \@firstoftwo
 \else \expandafter \@secondoftwo
 \fi
}%
\providecommand \@ifx [1]{%
 \ifx #1\expandafter \@firstoftwo
 \else \expandafter \@secondoftwo
 \fi
}%
\providecommand \natexlab [1]{#1}%
\providecommand \enquote  [1]{``#1''}%
\providecommand \bibnamefont  [1]{#1}%
\providecommand \bibfnamefont [1]{#1}%
\providecommand \citenamefont [1]{#1}%
\providecommand \href@noop [0]{\@secondoftwo}%
\providecommand \href [0]{\begingroup \@sanitize@url \@href}%
\providecommand \@href[1]{\@@startlink{#1}\@@href}%
\providecommand \@@href[1]{\endgroup#1\@@endlink}%
\providecommand \@sanitize@url [0]{\catcode `\\12\catcode `\$12\catcode
  `\&12\catcode `\#12\catcode `\^12\catcode `\_12\catcode `\%12\relax}%
\providecommand \@@startlink[1]{}%
\providecommand \@@endlink[0]{}%
\providecommand \url  [0]{\begingroup\@sanitize@url \@url }%
\providecommand \@url [1]{\endgroup\@href {#1}{\urlprefix }}%
\providecommand \urlprefix  [0]{URL }%
\providecommand \Eprint [0]{\href }%
\providecommand \doibase [0]{http://dx.doi.org/}%
\providecommand \selectlanguage [0]{\@gobble}%
\providecommand \bibinfo  [0]{\@secondoftwo}%
\providecommand \bibfield  [0]{\@secondoftwo}%
\providecommand \translation [1]{[#1]}%
\providecommand \BibitemOpen [0]{}%
\providecommand \bibitemStop [0]{}%
\providecommand \bibitemNoStop [0]{.\EOS\space}%
\providecommand \EOS [0]{\spacefactor3000\relax}%
\providecommand \BibitemShut  [1]{\csname bibitem#1\endcsname}%
\let\auto@bib@innerbib\@empty
%</preamble>
\bibitem [{\citenamefont {{Nayak}}\ \emph {et~al.}(2008)\citenamefont
  {{Nayak}}, \citenamefont {{Simon}}, \citenamefont {{Stern}}, \citenamefont
  {{Freedman}},\ and\ \citenamefont {{Das Sarma}}}]{Nayak08}%
  \BibitemOpen
  \bibfield  {author} {\bibinfo {author} {\bibfnamefont {C.}~\bibnamefont
  {{Nayak}}}, \bibinfo {author} {\bibfnamefont {S.~H.}\ \bibnamefont
  {{Simon}}}, \bibinfo {author} {\bibfnamefont {A.}~\bibnamefont {{Stern}}},
  \bibinfo {author} {\bibfnamefont {M.}~\bibnamefont {{Freedman}}}, \ and\
  \bibinfo {author} {\bibfnamefont {S.}~\bibnamefont {{Das Sarma}}},\
  }\bibfield  {title} {\enquote {\bibinfo {title} {{Non-Abelian anyons and
  topological quantum computation}},}\ }\href {\doibase
  10.1103/RevModPhys.80.1083} {\bibfield  {journal} {\bibinfo  {journal} {Rev.
  Mod. Phys.}\ }\textbf {\bibinfo {volume} {80}},\ \bibinfo {pages} {1083}
  (\bibinfo {year} {2008})},\ \Eprint {http://arxiv.org/abs/0707.1889}
  {arXiv:0707.1889} \BibitemShut {NoStop}%
\bibitem [{\citenamefont {{Das Sarma}}\ \emph {et~al.}(2015)\citenamefont {{Das
  Sarma}}, \citenamefont {{Freedman}},\ and\ \citenamefont
  {{Nayak}}}]{DasSarma2015}%
  \BibitemOpen
  \bibfield  {author} {\bibinfo {author} {\bibfnamefont {Sankar}\ \bibnamefont
  {{Das Sarma}}}, \bibinfo {author} {\bibfnamefont {Michael}\ \bibnamefont
  {{Freedman}}}, \ and\ \bibinfo {author} {\bibfnamefont {Chetan}\ \bibnamefont
  {{Nayak}}},\ }\bibfield  {title} {\enquote {\bibinfo {title} {{Majorana Zero
  Modes and Topological Quantum Computation}},}\ }\href@noop {} {\bibfield
  {journal} {\bibinfo  {journal} {arXiv e-prints}\ ,\ \bibinfo {eid}
  {arXiv:1501.02813}} (\bibinfo {year} {2015})},\ \Eprint
  {http://arxiv.org/abs/1501.02813} {arXiv:1501.02813 [cond-mat.str-el]}
  \BibitemShut {NoStop}%
\bibitem [{\citenamefont {{Kitaev}}(2001)}]{Kitaev01}%
  \BibitemOpen
  \bibfield  {author} {\bibinfo {author} {\bibfnamefont {A.~Y.}\ \bibnamefont
  {{Kitaev}}},\ }\bibfield  {title} {\enquote {\bibinfo {title} {Unpaired
  majorana fermions in quantum wires},}\ }\href {\doibase
  10.1070/1063-7869/44/10S/S29} {\bibfield  {journal} {\bibinfo  {journal}
  {Phys. Usp.}\ }\textbf {\bibinfo {volume} {44}},\ \bibinfo {pages} {131}
  (\bibinfo {year} {2001})},\ \Eprint {http://arxiv.org/abs/cond-mat/0010440}
  {cond-mat/0010440} \BibitemShut {NoStop}%
\bibitem [{\citenamefont {{Kitaev}}(2003)}]{Kitaev03}%
  \BibitemOpen
  \bibfield  {author} {\bibinfo {author} {\bibfnamefont {A.~Y.}\ \bibnamefont
  {{Kitaev}}},\ }\bibfield  {title} {\enquote {\bibinfo {title}
  {{Fault-tolerant quantum computation by anyons}},}\ }\href {\doibase
  10.1016/S0003-4916(02)00018-0} {\bibfield  {journal} {\bibinfo  {journal}
  {Ann. Phys.}\ }\textbf {\bibinfo {volume} {303}},\ \bibinfo {pages} {2}
  (\bibinfo {year} {2003})},\ \Eprint {http://arxiv.org/abs/quant-ph/9707021}
  {quant-ph/9707021} \BibitemShut {NoStop}%
\bibitem [{\citenamefont {{Alicea}}(2012)}]{Alicea12a}%
  \BibitemOpen
  \bibfield  {author} {\bibinfo {author} {\bibfnamefont {J.}~\bibnamefont
  {{Alicea}}},\ }\bibfield  {title} {\enquote {\bibinfo {title} {{New
  directions in the pursuit of Majorana fermions in solid state systems}},}\
  }\href {\doibase 10.1088/0034-4885/75/7/076501} {\bibfield  {journal}
  {\bibinfo  {journal} {Rep. Prog. Phys.}\ }\textbf {\bibinfo {volume} {75}},\
  \bibinfo {eid} {076501} (\bibinfo {year} {2012})},\ \Eprint
  {http://arxiv.org/abs/1202.1293} {arXiv:1202.1293} \BibitemShut {NoStop}%
\bibitem [{\citenamefont {Lutchyn}\ \emph {et~al.}(2018)\citenamefont
  {Lutchyn}, \citenamefont {Bakkers}, \citenamefont {Kouwenhoven},
  \citenamefont {Krogstrup}, \citenamefont {Marcus},\ and\ \citenamefont
  {Oreg}}]{Lutchyn17}%
  \BibitemOpen
  \bibfield  {author} {\bibinfo {author} {\bibfnamefont {R.~M.}\ \bibnamefont
  {Lutchyn}}, \bibinfo {author} {\bibfnamefont {E.~P. a.~M.}\ \bibnamefont
  {Bakkers}}, \bibinfo {author} {\bibfnamefont {L.~P.}\ \bibnamefont
  {Kouwenhoven}}, \bibinfo {author} {\bibfnamefont {P.}~\bibnamefont
  {Krogstrup}}, \bibinfo {author} {\bibfnamefont {C.~M.}\ \bibnamefont
  {Marcus}}, \ and\ \bibinfo {author} {\bibfnamefont {Y.}~\bibnamefont
  {Oreg}},\ }\bibfield  {title} {\enquote {\bibinfo {title} {Majorana zero
  modes in superconductor\textendash{}semiconductor heterostructures},}\ }\href
  {\doibase 10.1038/s41578-018-0003-1} {\bibfield  {journal} {\bibinfo
  {journal} {Nat. Rev. Mater.}\ }\textbf {\bibinfo {volume} {3}},\ \bibinfo
  {pages} {52} (\bibinfo {year} {2018})},\ \Eprint
  {http://arxiv.org/abs/arXiv:1707.04899} {arXiv:1707.04899} \BibitemShut
  {NoStop}%
\bibitem [{\citenamefont {Read}\ and\ \citenamefont {Green}(2000)}]{Read2000}%
  \BibitemOpen
  \bibfield  {author} {\bibinfo {author} {\bibfnamefont {N.}~\bibnamefont
  {Read}}\ and\ \bibinfo {author} {\bibfnamefont {Dmitry}\ \bibnamefont
  {Green}},\ }\bibfield  {title} {\enquote {\bibinfo {title} {Paired states of
  fermions in two dimensions with breaking of parity and time-reversal
  symmetries and the fractional quantum hall effect},}\ }\href {\doibase
  10.1103/PhysRevB.61.10267} {\bibfield  {journal} {\bibinfo  {journal} {Phys.
  Rev. B}\ }\textbf {\bibinfo {volume} {61}},\ \bibinfo {pages} {10267--10297}
  (\bibinfo {year} {2000})}\BibitemShut {NoStop}%
\bibitem [{\citenamefont {Ivanov}(2001)}]{Ivanov01}%
  \BibitemOpen
  \bibfield  {author} {\bibinfo {author} {\bibfnamefont {D.~A.}\ \bibnamefont
  {Ivanov}},\ }\bibfield  {title} {\enquote {\bibinfo {title} {Non-abelian
  statistics of half-quantum vortices in $\mathit{p}$-wave superconductors},}\
  }\href {\doibase 10.1103/PhysRevLett.86.268} {\bibfield  {journal} {\bibinfo
  {journal} {Phys. Rev. Lett.}\ }\textbf {\bibinfo {volume} {86}},\ \bibinfo
  {pages} {268--271} (\bibinfo {year} {2001})},\ \Eprint
  {http://arxiv.org/abs/cond-mat/0005069} {cond-mat/0005069} \BibitemShut
  {NoStop}%
\bibitem [{\citenamefont {{Lutchyn}}\ \emph {et~al.}(2010)\citenamefont
  {{Lutchyn}}, \citenamefont {{Sau}},\ and\ \citenamefont {{Das
  Sarma}}}]{Lutchyn10}%
  \BibitemOpen
  \bibfield  {author} {\bibinfo {author} {\bibfnamefont {R.~M.}\ \bibnamefont
  {{Lutchyn}}}, \bibinfo {author} {\bibfnamefont {J.~D.}\ \bibnamefont
  {{Sau}}}, \ and\ \bibinfo {author} {\bibfnamefont {S.}~\bibnamefont {{Das
  Sarma}}},\ }\bibfield  {title} {\enquote {\bibinfo {title} {{Majorana
  Fermions and a Topological Phase Transition in Semiconductor-Superconductor
  Heterostructures}},}\ }\href {\doibase 10.1103/PhysRevLett.105.077001}
  {\bibfield  {journal} {\bibinfo  {journal} {Phys. Rev. Lett.}\ }\textbf
  {\bibinfo {volume} {105}},\ \bibinfo {pages} {077001} (\bibinfo {year}
  {2010})},\ \Eprint {http://arxiv.org/abs/1002.4033} {arXiv:1002.4033}
  \BibitemShut {NoStop}%
\bibitem [{\citenamefont {{Oreg}}\ \emph {et~al.}(2010)\citenamefont {{Oreg}},
  \citenamefont {{Refael}},\ and\ \citenamefont {{von Oppen}}}]{Oreg10}%
  \BibitemOpen
  \bibfield  {author} {\bibinfo {author} {\bibfnamefont {Y.}~\bibnamefont
  {{Oreg}}}, \bibinfo {author} {\bibfnamefont {G.}~\bibnamefont {{Refael}}}, \
  and\ \bibinfo {author} {\bibfnamefont {F.}~\bibnamefont {{von Oppen}}},\
  }\bibfield  {title} {\enquote {\bibinfo {title} {{Helical Liquids and
  Majorana Bound States in Quantum Wires}},}\ }\href {\doibase
  10.1103/PhysRevLett.105.177002} {\bibfield  {journal} {\bibinfo  {journal}
  {Phys. Rev. Lett.}\ }\textbf {\bibinfo {volume} {105}},\ \bibinfo {pages}
  {177002} (\bibinfo {year} {2010})},\ \Eprint {http://arxiv.org/abs/1003.1145}
  {arXiv:1003.1145} \BibitemShut {NoStop}%
\bibitem [{\citenamefont {Mourik}\ \emph {et~al.}(2012)\citenamefont {Mourik},
  \citenamefont {Zuo}, \citenamefont {Frolov}, \citenamefont {Plissard},
  \citenamefont {Bakkers},\ and\ \citenamefont {Kouwenhoven}}]{Mourik12}%
  \BibitemOpen
  \bibfield  {author} {\bibinfo {author} {\bibfnamefont {V.}~\bibnamefont
  {Mourik}}, \bibinfo {author} {\bibfnamefont {K.}~\bibnamefont {Zuo}},
  \bibinfo {author} {\bibfnamefont {S.~M.}\ \bibnamefont {Frolov}}, \bibinfo
  {author} {\bibfnamefont {S.~R.}\ \bibnamefont {Plissard}}, \bibinfo {author}
  {\bibfnamefont {E.~P. A.~M.}\ \bibnamefont {Bakkers}}, \ and\ \bibinfo
  {author} {\bibfnamefont {L.~P.}\ \bibnamefont {Kouwenhoven}},\ }\bibfield
  {title} {\enquote {\bibinfo {title} {Signatures of {M}ajorana fermions in
  hybrid superconductor-semiconductor nanowire devices},}\ }\href {\doibase
  10.1126/science.1222360} {\bibfield  {journal} {\bibinfo  {journal}
  {Science}\ }\textbf {\bibinfo {volume} {336}},\ \bibinfo {pages} {1003}
  (\bibinfo {year} {2012})},\ \Eprint {http://arxiv.org/abs/arXiv:1204.2792}
  {arXiv:1204.2792} \BibitemShut {NoStop}%
\bibitem [{\citenamefont {Deng}\ \emph {et~al.}(2012)\citenamefont {Deng},
  \citenamefont {Yu}, \citenamefont {Huang}, \citenamefont {Larsson},
  \citenamefont {Caroff},\ and\ \citenamefont {Xu}}]{Deng12}%
  \BibitemOpen
  \bibfield  {author} {\bibinfo {author} {\bibfnamefont {M.~T.}\ \bibnamefont
  {Deng}}, \bibinfo {author} {\bibfnamefont {C.~L.}\ \bibnamefont {Yu}},
  \bibinfo {author} {\bibfnamefont {G.~Y.}\ \bibnamefont {Huang}}, \bibinfo
  {author} {\bibfnamefont {M.}~\bibnamefont {Larsson}}, \bibinfo {author}
  {\bibfnamefont {P.}~\bibnamefont {Caroff}}, \ and\ \bibinfo {author}
  {\bibfnamefont {H.~Q.}\ \bibnamefont {Xu}},\ }\bibfield  {title} {\enquote
  {\bibinfo {title} {Anomalous zero-bias conductance peak in a nb--insb
  nanowire--nb hybrid device},}\ }\href {\doibase 10.1021/nl303758w} {\bibfield
   {journal} {\bibinfo  {journal} {Nano Lett.}\ }\textbf {\bibinfo {volume}
  {12}},\ \bibinfo {pages} {6414} (\bibinfo {year} {2012})},\ \Eprint
  {http://arxiv.org/abs/arXiv:1204.4130} {arXiv:1204.4130} \BibitemShut
  {NoStop}%
\bibitem [{\citenamefont {Das}\ \emph {et~al.}(2012)\citenamefont {Das},
  \citenamefont {Ronen}, \citenamefont {Most}, \citenamefont {Oreg},
  \citenamefont {Heiblum},\ and\ \citenamefont {Shtrikman}}]{Das12}%
  \BibitemOpen
  \bibfield  {author} {\bibinfo {author} {\bibfnamefont {A.}~\bibnamefont
  {Das}}, \bibinfo {author} {\bibfnamefont {Y.}~\bibnamefont {Ronen}}, \bibinfo
  {author} {\bibfnamefont {Y.}~\bibnamefont {Most}}, \bibinfo {author}
  {\bibfnamefont {Y.}~\bibnamefont {Oreg}}, \bibinfo {author} {\bibfnamefont
  {M.}~\bibnamefont {Heiblum}}, \ and\ \bibinfo {author} {\bibfnamefont
  {H.}~\bibnamefont {Shtrikman}},\ }\bibfield  {title} {\enquote {\bibinfo
  {title} {Zero-bias peaks and splitting in an {Al-InAs} nanowire topological
  superconductor as a signature of {M}ajorana fermions},}\ }\href {\doibase
  10.1038/nphys2479} {\bibfield  {journal} {\bibinfo  {journal} {Nat. Phys.}\
  }\textbf {\bibinfo {volume} {8}},\ \bibinfo {pages} {887} (\bibinfo {year}
  {2012})},\ \Eprint {http://arxiv.org/abs/arXiv:1205.7073} {arXiv:1205.7073}
  \BibitemShut {NoStop}%
\bibitem [{\citenamefont {{Churchill}}\ \emph {et~al.}(2013)\citenamefont
  {{Churchill}}, \citenamefont {{Fatemi}}, \citenamefont {{Grove-Rasmussen}},
  \citenamefont {{Deng}}, \citenamefont {{Caroff}}, \citenamefont {{Xu}},\ and\
  \citenamefont {{Marcus}}}]{Churchill13}%
  \BibitemOpen
  \bibfield  {author} {\bibinfo {author} {\bibfnamefont {H.~O.~H.}\
  \bibnamefont {{Churchill}}}, \bibinfo {author} {\bibfnamefont
  {V.}~\bibnamefont {{Fatemi}}}, \bibinfo {author} {\bibfnamefont
  {K.}~\bibnamefont {{Grove-Rasmussen}}}, \bibinfo {author} {\bibfnamefont
  {M.~T.}\ \bibnamefont {{Deng}}}, \bibinfo {author} {\bibfnamefont
  {P.}~\bibnamefont {{Caroff}}}, \bibinfo {author} {\bibfnamefont {H.~Q.}\
  \bibnamefont {{Xu}}}, \ and\ \bibinfo {author} {\bibfnamefont {C.~M.}\
  \bibnamefont {{Marcus}}},\ }\bibfield  {title} {\enquote {\bibinfo {title}
  {{Superconductor-nanowire devices from tunneling to the multichannel regime:
  Zero-bias oscillations and magnetoconductance crossover}},}\ }\href {\doibase
  10.1103/PhysRevB.87.241401} {\bibfield  {journal} {\bibinfo  {journal} {Phys.
  Rev. B}\ }\textbf {\bibinfo {volume} {87}},\ \bibinfo {eid} {241401}
  (\bibinfo {year} {2013})},\ \Eprint {http://arxiv.org/abs/arXiv:1303.2407}
  {arXiv:1303.2407} \BibitemShut {NoStop}%
\bibitem [{\citenamefont {Finck}\ \emph {et~al.}(2013)\citenamefont {Finck},
  \citenamefont {Van~Harlingen}, \citenamefont {Mohseni}, \citenamefont
  {Jung},\ and\ \citenamefont {Li}}]{Finck12}%
  \BibitemOpen
  \bibfield  {author} {\bibinfo {author} {\bibfnamefont {A.~D.~K.}\
  \bibnamefont {Finck}}, \bibinfo {author} {\bibfnamefont {D.~J.}\ \bibnamefont
  {Van~Harlingen}}, \bibinfo {author} {\bibfnamefont {P.~K.}\ \bibnamefont
  {Mohseni}}, \bibinfo {author} {\bibfnamefont {K.}~\bibnamefont {Jung}}, \
  and\ \bibinfo {author} {\bibfnamefont {X.}~\bibnamefont {Li}},\ }\bibfield
  {title} {\enquote {\bibinfo {title} {{Anomalous Modulation of a Zero-Bias
  Peak in a Hybrid Nanowire-Superconductor Device}},}\ }\href {\doibase
  10.1103/PhysRevLett.110.126406} {\bibfield  {journal} {\bibinfo  {journal}
  {Phys. Rev. Lett.}\ }\textbf {\bibinfo {volume} {110}},\ \bibinfo {pages}
  {126406} (\bibinfo {year} {2013})},\ \Eprint
  {http://arxiv.org/abs/arXiv:1212.1101} {arXiv:1212.1101} \BibitemShut
  {NoStop}%
\bibitem [{\citenamefont {Deng}\ \emph {et~al.}(2016)\citenamefont {Deng},
  \citenamefont {Vaitiekėnas}, \citenamefont {Hansen}, \citenamefont {Danon},
  \citenamefont {Leijnse}, \citenamefont {Flensberg}, \citenamefont {Nygård},
  \citenamefont {Krogstrup},\ and\ \citenamefont {Marcus}}]{Deng16}%
  \BibitemOpen
  \bibfield  {author} {\bibinfo {author} {\bibfnamefont {M.~T.}\ \bibnamefont
  {Deng}}, \bibinfo {author} {\bibfnamefont {S.}~\bibnamefont {Vaitiekėnas}},
  \bibinfo {author} {\bibfnamefont {E.~B.}\ \bibnamefont {Hansen}}, \bibinfo
  {author} {\bibfnamefont {J.}~\bibnamefont {Danon}}, \bibinfo {author}
  {\bibfnamefont {M.}~\bibnamefont {Leijnse}}, \bibinfo {author} {\bibfnamefont
  {K.}~\bibnamefont {Flensberg}}, \bibinfo {author} {\bibfnamefont
  {J.}~\bibnamefont {Nygård}}, \bibinfo {author} {\bibfnamefont
  {P.}~\bibnamefont {Krogstrup}}, \ and\ \bibinfo {author} {\bibfnamefont
  {C.~M.}\ \bibnamefont {Marcus}},\ }\bibfield  {title} {\enquote {\bibinfo
  {title} {Majorana bound state in a coupled quantum-dot hybrid-nanowire
  system},}\ }\href {\doibase 10.1126/science.aaf3961} {\bibfield  {journal}
  {\bibinfo  {journal} {Science}\ }\textbf {\bibinfo {volume} {354}},\ \bibinfo
  {pages} {1557} (\bibinfo {year} {2016})},\ \Eprint
  {http://arxiv.org/abs/arXiv:1612.07989} {arXiv:1612.07989} \BibitemShut
  {NoStop}%
\bibitem [{\citenamefont {{Albrecht}}\ \emph {et~al.}(2016)\citenamefont
  {{Albrecht}}, \citenamefont {{Higginbotham}}, \citenamefont {{Madsen}},
  \citenamefont {{Kuemmeth}}, \citenamefont {{Jespersen}}, \citenamefont
  {{Nyg{\aa}rd}}, \citenamefont {{Krogstrup}},\ and\ \citenamefont
  {{Marcus}}}]{Albrecht16}%
  \BibitemOpen
  \bibfield  {author} {\bibinfo {author} {\bibfnamefont {S.~M.}\ \bibnamefont
  {{Albrecht}}}, \bibinfo {author} {\bibfnamefont {A.~P.}\ \bibnamefont
  {{Higginbotham}}}, \bibinfo {author} {\bibfnamefont {M.}~\bibnamefont
  {{Madsen}}}, \bibinfo {author} {\bibfnamefont {F.}~\bibnamefont
  {{Kuemmeth}}}, \bibinfo {author} {\bibfnamefont {T.~S.}\ \bibnamefont
  {{Jespersen}}}, \bibinfo {author} {\bibfnamefont {J.}~\bibnamefont
  {{Nyg{\aa}rd}}}, \bibinfo {author} {\bibfnamefont {P.}~\bibnamefont
  {{Krogstrup}}}, \ and\ \bibinfo {author} {\bibfnamefont {C.~M.}\ \bibnamefont
  {{Marcus}}},\ }\bibfield  {title} {\enquote {\bibinfo {title} {{Exponential
  protection of zero modes in Majorana islands}},}\ }\href {\doibase
  10.1038/nature17162} {\bibfield  {journal} {\bibinfo  {journal} {Nature}\
  }\textbf {\bibinfo {volume} {531}},\ \bibinfo {pages} {206} (\bibinfo {year}
  {2016})},\ \Eprint {http://arxiv.org/abs/arXiv:1603.03217} {arXiv:1603.03217}
  \BibitemShut {NoStop}%
\bibitem [{\citenamefont {{Nichele}}\ \emph {et~al.}(2017)\citenamefont
  {{Nichele}}, \citenamefont {{Drachmann}}, \citenamefont {{Whiticar}},
  \citenamefont {{O'Farrell}}, \citenamefont {{Suominen}}, \citenamefont
  {{Fornieri}}, \citenamefont {{Wang}}, \citenamefont {{Gardner}},
  \citenamefont {{Thomas}}, \citenamefont {{Hatke}}, \citenamefont
  {{Krogstrup}}, \citenamefont {{Manfra}}, \citenamefont {{Flensberg}},\ and\
  \citenamefont {{Marcus}}}]{Nichele17}%
  \BibitemOpen
  \bibfield  {author} {\bibinfo {author} {\bibfnamefont {F.}~\bibnamefont
  {{Nichele}}}, \bibinfo {author} {\bibfnamefont {A.~C.~C.}\ \bibnamefont
  {{Drachmann}}}, \bibinfo {author} {\bibfnamefont {A.~M.}\ \bibnamefont
  {{Whiticar}}}, \bibinfo {author} {\bibfnamefont {E.~C.~T.}\ \bibnamefont
  {{O'Farrell}}}, \bibinfo {author} {\bibfnamefont {H.~J.}\ \bibnamefont
  {{Suominen}}}, \bibinfo {author} {\bibfnamefont {A.}~\bibnamefont
  {{Fornieri}}}, \bibinfo {author} {\bibfnamefont {T.}~\bibnamefont {{Wang}}},
  \bibinfo {author} {\bibfnamefont {G.~C.}\ \bibnamefont {{Gardner}}}, \bibinfo
  {author} {\bibfnamefont {C.}~\bibnamefont {{Thomas}}}, \bibinfo {author}
  {\bibfnamefont {A.~T.}\ \bibnamefont {{Hatke}}}, \bibinfo {author}
  {\bibfnamefont {P.}~\bibnamefont {{Krogstrup}}}, \bibinfo {author}
  {\bibfnamefont {M.~J.}\ \bibnamefont {{Manfra}}}, \bibinfo {author}
  {\bibfnamefont {K.}~\bibnamefont {{Flensberg}}}, \ and\ \bibinfo {author}
  {\bibfnamefont {C.~M.}\ \bibnamefont {{Marcus}}},\ }\bibfield  {title}
  {\enquote {\bibinfo {title} {{Scaling of Majorana Zero-Bias Conductance
  Peaks}},}\ }\href {\doibase 10.1103/PhysRevLett.119.136803} {\bibfield
  {journal} {\bibinfo  {journal} {Phys. Rev. Lett.}\ }\textbf {\bibinfo
  {volume} {119}},\ \bibinfo {eid} {136803} (\bibinfo {year} {2017})},\ \Eprint
  {http://arxiv.org/abs/arXiv:1706.07033} {arXiv:1706.07033} \BibitemShut
  {NoStop}%
\bibitem [{\citenamefont {{Zhang}}\ \emph {et~al.}(2018)\citenamefont
  {{Zhang}}, \citenamefont {{Liu}}, \citenamefont {{Gazibegovic}},
  \citenamefont {{Xu}}, \citenamefont {{Logan}}, \citenamefont {{Wang}},
  \citenamefont {{van Loo}}, \citenamefont {{Bommer}}, \citenamefont {{de
  Moor}}, \citenamefont {{Car}}, \citenamefont {{Op Het Veld}}, \citenamefont
  {{van Veldhoven}}, \citenamefont {{Koelling}}, \citenamefont {{Verheijen}},
  \citenamefont {{Pendharkar}}, \citenamefont {{Pennachio}}, \citenamefont
  {{Shojaei}}, \citenamefont {{Lee}}, \citenamefont {{Palmstr{\o}m}},
  \citenamefont {{Bakkers}}, \citenamefont {{Sarma}},\ and\ \citenamefont
  {{Kouwenhoven}}}]{Zhang17}%
  \BibitemOpen
  \bibfield  {author} {\bibinfo {author} {\bibfnamefont {H.}~\bibnamefont
  {{Zhang}}}, \bibinfo {author} {\bibfnamefont {C.-X.}\ \bibnamefont {{Liu}}},
  \bibinfo {author} {\bibfnamefont {S.}~\bibnamefont {{Gazibegovic}}}, \bibinfo
  {author} {\bibfnamefont {D.}~\bibnamefont {{Xu}}}, \bibinfo {author}
  {\bibfnamefont {J.~A.}\ \bibnamefont {{Logan}}}, \bibinfo {author}
  {\bibfnamefont {G.}~\bibnamefont {{Wang}}}, \bibinfo {author} {\bibfnamefont
  {N.}~\bibnamefont {{van Loo}}}, \bibinfo {author} {\bibfnamefont {J.~D.~S.}\
  \bibnamefont {{Bommer}}}, \bibinfo {author} {\bibfnamefont {M.~W.~A.}\
  \bibnamefont {{de Moor}}}, \bibinfo {author} {\bibfnamefont {D.}~\bibnamefont
  {{Car}}}, \bibinfo {author} {\bibfnamefont {R.~L.~M.}\ \bibnamefont {{Op Het
  Veld}}}, \bibinfo {author} {\bibfnamefont {P.~J.}\ \bibnamefont {{van
  Veldhoven}}}, \bibinfo {author} {\bibfnamefont {S.}~\bibnamefont
  {{Koelling}}}, \bibinfo {author} {\bibfnamefont {M.~A.}\ \bibnamefont
  {{Verheijen}}}, \bibinfo {author} {\bibfnamefont {M.}~\bibnamefont
  {{Pendharkar}}}, \bibinfo {author} {\bibfnamefont {D.~J.}\ \bibnamefont
  {{Pennachio}}}, \bibinfo {author} {\bibfnamefont {B.}~\bibnamefont
  {{Shojaei}}}, \bibinfo {author} {\bibfnamefont {J.~S.}\ \bibnamefont
  {{Lee}}}, \bibinfo {author} {\bibfnamefont {C.~J.}\ \bibnamefont
  {{Palmstr{\o}m}}}, \bibinfo {author} {\bibfnamefont {E.~P.~A.~M.}\
  \bibnamefont {{Bakkers}}}, \bibinfo {author} {\bibfnamefont {S.~D.}\
  \bibnamefont {{Sarma}}}, \ and\ \bibinfo {author} {\bibfnamefont {L.~P.}\
  \bibnamefont {{Kouwenhoven}}},\ }\bibfield  {title} {\enquote {\bibinfo
  {title} {{Quantized Majorana conductance}},}\ }\href {\doibase
  10.1038/nature26142} {\bibfield  {journal} {\bibinfo  {journal} {Nature}\
  }\textbf {\bibinfo {volume} {556}},\ \bibinfo {pages} {74--79} (\bibinfo
  {year} {2018})},\ \Eprint {http://arxiv.org/abs/arXiv:1710.10701}
  {arXiv:1710.10701} \BibitemShut {NoStop}%
\bibitem [{\citenamefont {{Vaitiek{\.{e}}nas}}\ \emph
  {et~al.}(2018)\citenamefont {{Vaitiek{\.{e}}nas}}, \citenamefont {{Deng}},
  \citenamefont {{Krogstrup}},\ and\ \citenamefont {{Marcus}}}]{Vaitiekenas18}%
  \BibitemOpen
  \bibfield  {author} {\bibinfo {author} {\bibfnamefont {S.}~\bibnamefont
  {{Vaitiek{\.{e}}nas}}}, \bibinfo {author} {\bibfnamefont {M.~T.}\
  \bibnamefont {{Deng}}}, \bibinfo {author} {\bibfnamefont {P.}~\bibnamefont
  {{Krogstrup}}}, \ and\ \bibinfo {author} {\bibfnamefont {C.~M.}\ \bibnamefont
  {{Marcus}}},\ }\bibfield  {title} {\enquote {\bibinfo {title} {{Flux-induced
  Majorana modes in full-shell nanowires}},}\ }\href@noop {} {\bibfield
  {journal} {\bibinfo  {journal} {arXiv e-prints}\ } (\bibinfo {year}
  {2018})},\ \Eprint {http://arxiv.org/abs/arXiv:1809.05513} {arXiv:1809.05513}
  \BibitemShut {NoStop}%
\bibitem [{\citenamefont {Sau}\ \emph {et~al.}(2010)\citenamefont {Sau},
  \citenamefont {Tewari},\ and\ \citenamefont {Das~Sarma}}]{Sau2010a}%
  \BibitemOpen
  \bibfield  {author} {\bibinfo {author} {\bibfnamefont {Jay~D.}\ \bibnamefont
  {Sau}}, \bibinfo {author} {\bibfnamefont {Sumanta}\ \bibnamefont {Tewari}}, \
  and\ \bibinfo {author} {\bibfnamefont {S.}~\bibnamefont {Das~Sarma}},\
  }\bibfield  {title} {\enquote {\bibinfo {title} {Universal quantum
  computation in a semiconductor quantum wire network},}\ }\href {\doibase
  10.1103/PhysRevA.82.052322} {\bibfield  {journal} {\bibinfo  {journal} {Phys.
  Rev. A}\ }\textbf {\bibinfo {volume} {82}},\ \bibinfo {pages} {052322}
  (\bibinfo {year} {2010})}\BibitemShut {NoStop}%
\bibitem [{\citenamefont {{Hassler}}\ \emph {et~al.}(2011)\citenamefont
  {{Hassler}}, \citenamefont {{Akhmerov}},\ and\ \citenamefont
  {{Beenakker}}}]{Hassler11}%
  \BibitemOpen
  \bibfield  {author} {\bibinfo {author} {\bibfnamefont {F.}~\bibnamefont
  {{Hassler}}}, \bibinfo {author} {\bibfnamefont {A.~R.}\ \bibnamefont
  {{Akhmerov}}}, \ and\ \bibinfo {author} {\bibfnamefont {C.~W.~J.}\
  \bibnamefont {{Beenakker}}},\ }\bibfield  {title} {\enquote {\bibinfo {title}
  {{The top-transmon: a hybrid superconducting qubit for parity-protected
  quantum computation}},}\ }\href {\doibase 10.1088/1367-2630/13/9/095004}
  {\bibfield  {journal} {\bibinfo  {journal} {New J. Phys.}\ }\textbf {\bibinfo
  {volume} {13}},\ \bibinfo {eid} {095004} (\bibinfo {year} {2011})},\ \Eprint
  {http://arxiv.org/abs/arXiv:1105.0315} {arXiv:1105.0315} \BibitemShut
  {NoStop}%
\bibitem [{\citenamefont {{van Heck}}\ \emph {et~al.}(2012)\citenamefont {{van
  Heck}}, \citenamefont {{Akhmerov}}, \citenamefont {{Hassler}}, \citenamefont
  {{Burrello}},\ and\ \citenamefont {{Beenakker}}}]{vanHeck11}%
  \BibitemOpen
  \bibfield  {author} {\bibinfo {author} {\bibfnamefont {B.}~\bibnamefont {{van
  Heck}}}, \bibinfo {author} {\bibfnamefont {A.~R.}\ \bibnamefont
  {{Akhmerov}}}, \bibinfo {author} {\bibfnamefont {F.}~\bibnamefont
  {{Hassler}}}, \bibinfo {author} {\bibfnamefont {M.}~\bibnamefont
  {{Burrello}}}, \ and\ \bibinfo {author} {\bibfnamefont {C.~W.~J.}\
  \bibnamefont {{Beenakker}}},\ }\bibfield  {title} {\enquote {\bibinfo {title}
  {{Coulomb-assisted braiding of Majorana fermions in a Josephson junction
  array}},}\ }\href {\doibase 10.1088/1367-2630/14/3/035019} {\bibfield
  {journal} {\bibinfo  {journal} {New J. Phys.}\ }\textbf {\bibinfo {volume}
  {14}},\ \bibinfo {eid} {035019} (\bibinfo {year} {2012})},\ \Eprint
  {http://arxiv.org/abs/arXiv:1111.6001} {arXiv:1111.6001} \BibitemShut
  {NoStop}%
\bibitem [{\citenamefont {{Hyart}}\ \emph {et~al.}(2013)\citenamefont
  {{Hyart}}, \citenamefont {{van Heck}}, \citenamefont {{Fulga}}, \citenamefont
  {{Burrello}}, \citenamefont {{Akhmerov}},\ and\ \citenamefont
  {{Beenakker}}}]{Hyart13}%
  \BibitemOpen
  \bibfield  {author} {\bibinfo {author} {\bibfnamefont {T.}~\bibnamefont
  {{Hyart}}}, \bibinfo {author} {\bibfnamefont {B.}~\bibnamefont {{van Heck}}},
  \bibinfo {author} {\bibfnamefont {I.~C.}\ \bibnamefont {{Fulga}}}, \bibinfo
  {author} {\bibfnamefont {M.}~\bibnamefont {{Burrello}}}, \bibinfo {author}
  {\bibfnamefont {A.~R.}\ \bibnamefont {{Akhmerov}}}, \ and\ \bibinfo {author}
  {\bibfnamefont {C.~W.~J.}\ \bibnamefont {{Beenakker}}},\ }\bibfield  {title}
  {\enquote {\bibinfo {title} {{Flux-controlled quantum computation with
  Majorana fermions}},}\ }\href {\doibase 10.1103/PhysRevB.88.035121}
  {\bibfield  {journal} {\bibinfo  {journal} {Phys. Rev. B}\ }\textbf {\bibinfo
  {volume} {88}},\ \bibinfo {eid} {035121} (\bibinfo {year} {2013})},\ \Eprint
  {http://arxiv.org/abs/arXiv:1303.4379} {arXiv:1303.4379} \BibitemShut
  {NoStop}%
\bibitem [{\citenamefont {{Aasen}}\ \emph {et~al.}(2016)\citenamefont
  {{Aasen}}, \citenamefont {{Hell}}, \citenamefont {{Mishmash}}, \citenamefont
  {{Higginbotham}}, \citenamefont {{Danon}}, \citenamefont {{Leijnse}},
  \citenamefont {{Jespersen}}, \citenamefont {{Folk}}, \citenamefont
  {{Marcus}}, \citenamefont {{Flensberg}},\ and\ \citenamefont
  {{Alicea}}}]{Aasen16}%
  \BibitemOpen
  \bibfield  {author} {\bibinfo {author} {\bibfnamefont {D.}~\bibnamefont
  {{Aasen}}}, \bibinfo {author} {\bibfnamefont {M.}~\bibnamefont {{Hell}}},
  \bibinfo {author} {\bibfnamefont {R.~V.}\ \bibnamefont {{Mishmash}}},
  \bibinfo {author} {\bibfnamefont {A.}~\bibnamefont {{Higginbotham}}},
  \bibinfo {author} {\bibfnamefont {J.}~\bibnamefont {{Danon}}}, \bibinfo
  {author} {\bibfnamefont {M.}~\bibnamefont {{Leijnse}}}, \bibinfo {author}
  {\bibfnamefont {T.~S.}\ \bibnamefont {{Jespersen}}}, \bibinfo {author}
  {\bibfnamefont {J.~A.}\ \bibnamefont {{Folk}}}, \bibinfo {author}
  {\bibfnamefont {C.~M.}\ \bibnamefont {{Marcus}}}, \bibinfo {author}
  {\bibfnamefont {K.}~\bibnamefont {{Flensberg}}}, \ and\ \bibinfo {author}
  {\bibfnamefont {J.}~\bibnamefont {{Alicea}}},\ }\bibfield  {title} {\enquote
  {\bibinfo {title} {{Milestones Toward Majorana-Based Quantum Computing}},}\
  }\href {\doibase 10.1103/PhysRevX.6.031016} {\bibfield  {journal} {\bibinfo
  {journal} {Phys. Rev. X}\ }\textbf {\bibinfo {volume} {6}},\ \bibinfo {eid}
  {031016} (\bibinfo {year} {2016})},\ \Eprint
  {http://arxiv.org/abs/arXiv:1511.05153} {arXiv:1511.05153} \BibitemShut
  {NoStop}%
\bibitem [{\citenamefont {Karzig}\ \emph {et~al.}(2017)\citenamefont {Karzig},
  \citenamefont {Knapp}, \citenamefont {Lutchyn}, \citenamefont {Bonderson},
  \citenamefont {Hastings}, \citenamefont {Nayak}, \citenamefont {Alicea},
  \citenamefont {Flensberg}, \citenamefont {Plugge}, \citenamefont {Oreg},
  \citenamefont {Marcus},\ and\ \citenamefont {Freedman}}]{Karzig17}%
  \BibitemOpen
  \bibfield  {author} {\bibinfo {author} {\bibfnamefont {Torsten}\ \bibnamefont
  {Karzig}}, \bibinfo {author} {\bibfnamefont {Christina}\ \bibnamefont
  {Knapp}}, \bibinfo {author} {\bibfnamefont {Roman~M.}\ \bibnamefont
  {Lutchyn}}, \bibinfo {author} {\bibfnamefont {Parsa}\ \bibnamefont
  {Bonderson}}, \bibinfo {author} {\bibfnamefont {Matthew~B.}\ \bibnamefont
  {Hastings}}, \bibinfo {author} {\bibfnamefont {Chetan}\ \bibnamefont
  {Nayak}}, \bibinfo {author} {\bibfnamefont {Jason}\ \bibnamefont {Alicea}},
  \bibinfo {author} {\bibfnamefont {Karsten}\ \bibnamefont {Flensberg}},
  \bibinfo {author} {\bibfnamefont {Stephan}\ \bibnamefont {Plugge}}, \bibinfo
  {author} {\bibfnamefont {Yuval}\ \bibnamefont {Oreg}}, \bibinfo {author}
  {\bibfnamefont {Charles~M.}\ \bibnamefont {Marcus}}, \ and\ \bibinfo {author}
  {\bibfnamefont {Michael~H.}\ \bibnamefont {Freedman}},\ }\bibfield  {title}
  {\enquote {\bibinfo {title} {Scalable designs for
  quasiparticle-poisoning-protected topological quantum computation with
  {{Majorana}} zero modes},}\ }\href {\doibase 10.1103/PhysRevB.95.235305}
  {\bibfield  {journal} {\bibinfo  {journal} {Phys. Rev. B}\ }\textbf {\bibinfo
  {volume} {95}},\ \bibinfo {pages} {235305} (\bibinfo {year} {2017})},\
  \Eprint {http://arxiv.org/abs/arXiv:1610.05289} {arXiv:1610.05289}
  \BibitemShut {NoStop}%
\bibitem [{\citenamefont {{Plugge}}\ \emph {et~al.}(2017)\citenamefont
  {{Plugge}}, \citenamefont {{Rasmussen}}, \citenamefont {{Egger}},\ and\
  \citenamefont {{Flensberg}}}]{Plugge17}%
  \BibitemOpen
  \bibfield  {author} {\bibinfo {author} {\bibfnamefont {S.}~\bibnamefont
  {{Plugge}}}, \bibinfo {author} {\bibfnamefont {A.}~\bibnamefont
  {{Rasmussen}}}, \bibinfo {author} {\bibfnamefont {R.}~\bibnamefont
  {{Egger}}}, \ and\ \bibinfo {author} {\bibfnamefont {K.}~\bibnamefont
  {{Flensberg}}},\ }\bibfield  {title} {\enquote {\bibinfo {title} {{Majorana
  box qubits}},}\ }\href {\doibase 10.1088/1367-2630/aa54e1} {\bibfield
  {journal} {\bibinfo  {journal} {New J. Phys.}\ }\textbf {\bibinfo {volume}
  {19}},\ \bibinfo {eid} {012001} (\bibinfo {year} {2017})},\ \Eprint
  {http://arxiv.org/abs/arXiv:1609.01697} {arXiv:1609.01697} \BibitemShut
  {NoStop}%
\bibitem [{\citenamefont {{Vijay}}\ and\ \citenamefont {{Fu}}(2016)}]{Vijay16}%
  \BibitemOpen
  \bibfield  {author} {\bibinfo {author} {\bibfnamefont {S.}~\bibnamefont
  {{Vijay}}}\ and\ \bibinfo {author} {\bibfnamefont {L.}~\bibnamefont {{Fu}}},\
  }\bibfield  {title} {\enquote {\bibinfo {title} {{Teleportation-based quantum
  information processing with Majorana zero modes}},}\ }\href {\doibase
  10.1103/PhysRevB.94.235446} {\bibfield  {journal} {\bibinfo  {journal} {Phys.
  Rev. B}\ }\textbf {\bibinfo {volume} {94}},\ \bibinfo {eid} {235446}
  (\bibinfo {year} {2016})},\ \Eprint {http://arxiv.org/abs/arXiv:1609.00950}
  {arXiv:1609.00950} \BibitemShut {NoStop}%
\bibitem [{\citenamefont {Bonderson}\ \emph {et~al.}(2008)\citenamefont
  {Bonderson}, \citenamefont {Freedman},\ and\ \citenamefont
  {Nayak}}]{Bonderson08b}%
  \BibitemOpen
  \bibfield  {author} {\bibinfo {author} {\bibfnamefont {P.}~\bibnamefont
  {Bonderson}}, \bibinfo {author} {\bibfnamefont {M.}~\bibnamefont {Freedman}},
  \ and\ \bibinfo {author} {\bibfnamefont {C.}~\bibnamefont {Nayak}},\
  }\bibfield  {title} {\enquote {\bibinfo {title} {Measurement-only topological
  quantum computation},}\ }\href {\doibase 10.1103/PhysRevLett.101.010501}
  {\bibfield  {journal} {\bibinfo  {journal} {Phys. Rev. Lett.}\ }\textbf
  {\bibinfo {volume} {101}},\ \bibinfo {pages} {010501} (\bibinfo {year}
  {2008})},\ \Eprint {http://arxiv.org/abs/arXiv:0802.0279} {arXiv:0802.0279}
  \BibitemShut {NoStop}%
\bibitem [{\citenamefont {Bonderson}\ \emph {et~al.}(2009)\citenamefont
  {Bonderson}, \citenamefont {Freedman},\ and\ \citenamefont
  {Nayak}}]{Bonderson08c}%
  \BibitemOpen
  \bibfield  {author} {\bibinfo {author} {\bibfnamefont {P.}~\bibnamefont
  {Bonderson}}, \bibinfo {author} {\bibfnamefont {M.}~\bibnamefont {Freedman}},
  \ and\ \bibinfo {author} {\bibfnamefont {C.}~\bibnamefont {Nayak}},\
  }\bibfield  {title} {\enquote {\bibinfo {title} {Measurement-only topological
  quantum computation via anyonic interferometry},}\ }\href {\doibase
  10.1016/j.aop.2008.09.009} {\bibfield  {journal} {\bibinfo  {journal} {Ann.
  Phys.}\ }\textbf {\bibinfo {volume} {324}},\ \bibinfo {pages} {787} (\bibinfo
  {year} {2009})},\ \Eprint {http://arxiv.org/abs/arXiv:0808.1933}
  {arXiv:0808.1933} \BibitemShut {NoStop}%
\bibitem [{\citenamefont {{Zheng}}\ \emph {et~al.}(2016)\citenamefont
  {{Zheng}}, \citenamefont {{Dua}},\ and\ \citenamefont {{Jiang}}}]{Zhang16}%
  \BibitemOpen
  \bibfield  {author} {\bibinfo {author} {\bibfnamefont {Huaixiu}\ \bibnamefont
  {{Zheng}}}, \bibinfo {author} {\bibfnamefont {Arpit}\ \bibnamefont {{Dua}}},
  \ and\ \bibinfo {author} {\bibfnamefont {Liang}\ \bibnamefont {{Jiang}}},\
  }\bibfield  {title} {\enquote {\bibinfo {title} {{Measurement-only
  topological quantum computation without forced measurements}},}\ }\href
  {\doibase 10.1088/1367-2630/aa50bb} {\bibfield  {journal} {\bibinfo
  {journal} {New Journal of Physics}\ }\textbf {\bibinfo {volume} {18}},\
  \bibinfo {eid} {123027} (\bibinfo {year} {2016})},\ \Eprint
  {http://arxiv.org/abs/arXiv:1607.07475} {arXiv:1607.07475} \BibitemShut
  {NoStop}%
\bibitem [{\citenamefont {{Tran}}\ \emph {et~al.}(2019)\citenamefont {{Tran}},
  \citenamefont {{Bocharov}}, \citenamefont {{Bauer}},\ and\ \citenamefont
  {{Bonderson}}}]{Tran19}%
  \BibitemOpen
  \bibfield  {author} {\bibinfo {author} {\bibfnamefont {Alan}\ \bibnamefont
  {{Tran}}}, \bibinfo {author} {\bibfnamefont {Alex}\ \bibnamefont
  {{Bocharov}}}, \bibinfo {author} {\bibfnamefont {Bela}\ \bibnamefont
  {{Bauer}}}, \ and\ \bibinfo {author} {\bibfnamefont {Parsa}\ \bibnamefont
  {{Bonderson}}},\ }\bibfield  {title} {\enquote {\bibinfo {title} {{Optimizing
  Clifford gate generation for measurement-only topological quantum computation
  with Majorana zero modes}},}\ }\href@noop {} {\bibfield  {journal} {\bibinfo
  {journal} {arXiv e-prints}\ } (\bibinfo {year} {2019})},\ \Eprint
  {http://arxiv.org/abs/arXiv:1909.03002} {arXiv:1909.03002} \BibitemShut
  {NoStop}%
\bibitem [{\citenamefont {{Weda Bomantara}}\ and\ \citenamefont
  {{Gong}}(2019)}]{Bomantara19}%
  \BibitemOpen
  \bibfield  {author} {\bibinfo {author} {\bibfnamefont {Raditya}\ \bibnamefont
  {{Weda Bomantara}}}\ and\ \bibinfo {author} {\bibfnamefont {Jiangbin}\
  \bibnamefont {{Gong}}},\ }\bibfield  {title} {\enquote {\bibinfo {title}
  {{Measurement-only quantum computation with Majorana corner modes}},}\
  }\href@noop {} {\  (\bibinfo {year} {2019})},\ \Eprint
  {http://arxiv.org/abs/arXiv:1904.03161} {arXiv:1904.03161} \BibitemShut
  {NoStop}%
\bibitem [{\citenamefont {{Alicea}}\ \emph {et~al.}(2011)\citenamefont
  {{Alicea}}, \citenamefont {{Oreg}}, \citenamefont {{Refael}}, \citenamefont
  {{von Oppen}},\ and\ \citenamefont {{Fisher}}}]{Alicea11}%
  \BibitemOpen
  \bibfield  {author} {\bibinfo {author} {\bibfnamefont {J.}~\bibnamefont
  {{Alicea}}}, \bibinfo {author} {\bibfnamefont {Y.}~\bibnamefont {{Oreg}}},
  \bibinfo {author} {\bibfnamefont {G.}~\bibnamefont {{Refael}}}, \bibinfo
  {author} {\bibfnamefont {F.}~\bibnamefont {{von Oppen}}}, \ and\ \bibinfo
  {author} {\bibfnamefont {M.~P.~A.}\ \bibnamefont {{Fisher}}},\ }\bibfield
  {title} {\enquote {\bibinfo {title} {{Non-Abelian statistics and topological
  quantum information processing in 1D wire networks}},}\ }\href {\doibase
  10.1038/nphys1915} {\bibfield  {journal} {\bibinfo  {journal} {Nature
  Physics}\ }\textbf {\bibinfo {volume} {7}},\ \bibinfo {pages} {412--417}
  (\bibinfo {year} {2011})},\ \Eprint {http://arxiv.org/abs/1006.4395}
  {arXiv:1006.4395} \BibitemShut {NoStop}%
\bibitem [{\citenamefont {{Bauer}}\ \emph {et~al.}(2018)\citenamefont
  {{Bauer}}, \citenamefont {{Karzig}}, \citenamefont {{Mishmash}},
  \citenamefont {{Antipov}},\ and\ \citenamefont {{Alicea}}}]{Bauer18}%
  \BibitemOpen
  \bibfield  {author} {\bibinfo {author} {\bibfnamefont {Bela}\ \bibnamefont
  {{Bauer}}}, \bibinfo {author} {\bibfnamefont {Torsten}\ \bibnamefont
  {{Karzig}}}, \bibinfo {author} {\bibfnamefont {Ryan}\ \bibnamefont
  {{Mishmash}}}, \bibinfo {author} {\bibfnamefont {Andrey}\ \bibnamefont
  {{Antipov}}}, \ and\ \bibinfo {author} {\bibfnamefont {Jason}\ \bibnamefont
  {{Alicea}}},\ }\bibfield  {title} {\enquote {\bibinfo {title} {{Dynamics of
  Majorana-based qubits operated with an array of tunable gates}},}\ }\href
  {\doibase 10.21468/SciPostPhys.5.1.004} {\bibfield  {journal} {\bibinfo
  {journal} {SciPost Physics}\ }\textbf {\bibinfo {volume} {5}},\ \bibinfo
  {eid} {004} (\bibinfo {year} {2018})},\ \Eprint
  {http://arxiv.org/abs/arXiv:1803.05451} {arXiv:1803.05451} \BibitemShut
  {NoStop}%
\bibitem [{\citenamefont {Knapp}\ \emph {et~al.}(2016)\citenamefont {Knapp},
  \citenamefont {Zaletel}, \citenamefont {Liu}, \citenamefont {Cheng},
  \citenamefont {Bonderson},\ and\ \citenamefont {Nayak}}]{Knapp16}%
  \BibitemOpen
  \bibfield  {author} {\bibinfo {author} {\bibfnamefont {Christina}\
  \bibnamefont {Knapp}}, \bibinfo {author} {\bibfnamefont {Michael}\
  \bibnamefont {Zaletel}}, \bibinfo {author} {\bibfnamefont {Dong~E.}\
  \bibnamefont {Liu}}, \bibinfo {author} {\bibfnamefont {Meng}\ \bibnamefont
  {Cheng}}, \bibinfo {author} {\bibfnamefont {Parsa}\ \bibnamefont
  {Bonderson}}, \ and\ \bibinfo {author} {\bibfnamefont {Chetan}\ \bibnamefont
  {Nayak}},\ }\bibfield  {title} {\enquote {\bibinfo {title} {The nature and
  correction of diabatic errors in anyon braiding},}\ }\href {\doibase
  10.1103/PhysRevX.6.041003} {\bibfield  {journal} {\bibinfo  {journal} {Phys.
  Rev. X}\ }\textbf {\bibinfo {volume} {6}},\ \bibinfo {pages} {041003}
  (\bibinfo {year} {2016})},\ \Eprint {http://arxiv.org/abs/arXiv:1601.05790}
  {arXiv:1601.05790} \BibitemShut {NoStop}%
\bibitem [{\citenamefont {Ortiz}\ \emph {et~al.}(2014)\citenamefont {Ortiz},
  \citenamefont {Dukelsky}, \citenamefont {Cobanera}, \citenamefont {Esebbag},\
  and\ \citenamefont {Beenakker}}]{Ortiz14}%
  \BibitemOpen
  \bibfield  {author} {\bibinfo {author} {\bibfnamefont {Gerardo}\ \bibnamefont
  {Ortiz}}, \bibinfo {author} {\bibfnamefont {Jorge}\ \bibnamefont {Dukelsky}},
  \bibinfo {author} {\bibfnamefont {Emilio}\ \bibnamefont {Cobanera}}, \bibinfo
  {author} {\bibfnamefont {Carlos}\ \bibnamefont {Esebbag}}, \ and\ \bibinfo
  {author} {\bibfnamefont {Carlo}\ \bibnamefont {Beenakker}},\ }\bibfield
  {title} {\enquote {\bibinfo {title} {Many-body characterization of
  particle-conserving topological superfluids},}\ }\href {\doibase
  10.1103/PhysRevLett.113.267002} {\bibfield  {journal} {\bibinfo  {journal}
  {Phys. Rev. Lett.}\ }\textbf {\bibinfo {volume} {113}},\ \bibinfo {pages}
  {267002} (\bibinfo {year} {2014})}\BibitemShut {NoStop}%
\bibitem [{\citenamefont {Ortiz}\ and\ \citenamefont
  {Cobanera}(2016)}]{Ortiz16}%
  \BibitemOpen
  \bibfield  {author} {\bibinfo {author} {\bibfnamefont {Gerardo}\ \bibnamefont
  {Ortiz}}\ and\ \bibinfo {author} {\bibfnamefont {Emilio}\ \bibnamefont
  {Cobanera}},\ }\bibfield  {title} {\enquote {\bibinfo {title} {What is a
  particle-conserving topological superfluid? the fate of majorana modes beyond
  mean-field theory},}\ }\href {\doibase
  https://doi.org/10.1016/j.aop.2016.05.020} {\bibfield  {journal} {\bibinfo
  {journal} {Annals of Physics}\ }\textbf {\bibinfo {volume} {372}},\ \bibinfo
  {pages} {357 -- 374} (\bibinfo {year} {2016})}\BibitemShut {NoStop}%
\bibitem [{\citenamefont {Wang}\ \emph {et~al.}(2017)\citenamefont {Wang},
  \citenamefont {Xu}, \citenamefont {Pu},\ and\ \citenamefont
  {Hazzard}}]{Wang17}%
  \BibitemOpen
  \bibfield  {author} {\bibinfo {author} {\bibfnamefont {Zhiyuan}\ \bibnamefont
  {Wang}}, \bibinfo {author} {\bibfnamefont {Youjiang}\ \bibnamefont {Xu}},
  \bibinfo {author} {\bibfnamefont {Han}\ \bibnamefont {Pu}}, \ and\ \bibinfo
  {author} {\bibfnamefont {Kaden R.~A.}\ \bibnamefont {Hazzard}},\ }\bibfield
  {title} {\enquote {\bibinfo {title} {Number-conserving interacting fermion
  models with exact topological superconducting ground states},}\ }\href
  {\doibase 10.1103/PhysRevB.96.115110} {\bibfield  {journal} {\bibinfo
  {journal} {Phys. Rev. B}\ }\textbf {\bibinfo {volume} {96}},\ \bibinfo
  {pages} {115110} (\bibinfo {year} {2017})}\BibitemShut {NoStop}%
\bibitem [{\citenamefont {Wang}\ and\ \citenamefont {Hazzard}(2018)}]{Wang18}%
  \BibitemOpen
  \bibfield  {author} {\bibinfo {author} {\bibfnamefont {Zhiyuan}\ \bibnamefont
  {Wang}}\ and\ \bibinfo {author} {\bibfnamefont {Kaden R.~A.}\ \bibnamefont
  {Hazzard}},\ }\bibfield  {title} {\enquote {\bibinfo {title} {Analytic ground
  state wave functions of mean-field ${p}_{x}+i{p}_{y}$ superconductors with
  vortices and boundaries},}\ }\href {\doibase 10.1103/PhysRevB.97.104501}
  {\bibfield  {journal} {\bibinfo  {journal} {Phys. Rev. B}\ }\textbf {\bibinfo
  {volume} {97}},\ \bibinfo {pages} {104501} (\bibinfo {year}
  {2018})}\BibitemShut {NoStop}%
\bibitem [{\citenamefont {{Lin}}\ and\ \citenamefont
  {{Leggett}}(2017)}]{Lin17}%
  \BibitemOpen
  \bibfield  {author} {\bibinfo {author} {\bibfnamefont {Yiruo}\ \bibnamefont
  {{Lin}}}\ and\ \bibinfo {author} {\bibfnamefont {Anthony~J.}\ \bibnamefont
  {{Leggett}}},\ }\bibfield  {title} {\enquote {\bibinfo {title} {{Effect of
  Particle Number Conservation on the Berry Phase Resulting from Transport of a
  Bound Quasiparticle around a Superfluid Vortex}},}\ }\href@noop {} {\
  (\bibinfo {year} {2017})},\ \Eprint {http://arxiv.org/abs/arXiv:1708.02578}
  {arXiv:1708.02578} \BibitemShut {NoStop}%
\bibitem [{\citenamefont {{Lin}}\ and\ \citenamefont
  {{Leggett}}(2018)}]{Lin18}%
  \BibitemOpen
  \bibfield  {author} {\bibinfo {author} {\bibfnamefont {Yiruo}\ \bibnamefont
  {{Lin}}}\ and\ \bibinfo {author} {\bibfnamefont {Anthony~J.}\ \bibnamefont
  {{Leggett}}},\ }\bibfield  {title} {\enquote {\bibinfo {title} {{Towards a
  Particle-Number Conserving Theory of Majorana Zero Modes in p+ip
  Superfluids}},}\ }\href@noop {} {\  (\bibinfo {year} {2018})},\ \Eprint
  {http://arxiv.org/abs/arXiv:1803.08003} {arXiv:1803.08003} \BibitemShut
  {NoStop}%
\bibitem [{\citenamefont {Sau}\ \emph {et~al.}(2011)\citenamefont {Sau},
  \citenamefont {Halperin}, \citenamefont {Flensberg},\ and\ \citenamefont
  {Das~Sarma}}]{Sau11}%
  \BibitemOpen
  \bibfield  {author} {\bibinfo {author} {\bibfnamefont {Jay~D.}\ \bibnamefont
  {Sau}}, \bibinfo {author} {\bibfnamefont {B.~I.}\ \bibnamefont {Halperin}},
  \bibinfo {author} {\bibfnamefont {K.}~\bibnamefont {Flensberg}}, \ and\
  \bibinfo {author} {\bibfnamefont {S.}~\bibnamefont {Das~Sarma}},\ }\bibfield
  {title} {\enquote {\bibinfo {title} {Number conserving theory for
  topologically protected degeneracy in one-dimensional fermions},}\ }\href
  {\doibase 10.1103/physrevb.84.144509} {\bibfield  {journal} {\bibinfo
  {journal} {Physical Review B}\ }\textbf {\bibinfo {volume} {84}} (\bibinfo
  {year} {2011}),\ 10.1103/physrevb.84.144509},\ \Eprint
  {http://arxiv.org/abs/arXiv:1106.4014} {arXiv:1106.4014} \BibitemShut
  {NoStop}%
\bibitem [{\citenamefont {{Fidkowski}}\ \emph {et~al.}(2011)\citenamefont
  {{Fidkowski}}, \citenamefont {{Lutchyn}}, \citenamefont {{Nayak}},\ and\
  \citenamefont {{Fisher}}}]{Fidkowski11}%
  \BibitemOpen
  \bibfield  {author} {\bibinfo {author} {\bibfnamefont {Lukasz}\ \bibnamefont
  {{Fidkowski}}}, \bibinfo {author} {\bibfnamefont {Roman~M.}\ \bibnamefont
  {{Lutchyn}}}, \bibinfo {author} {\bibfnamefont {Chetan}\ \bibnamefont
  {{Nayak}}}, \ and\ \bibinfo {author} {\bibfnamefont {Matthew P.~A.}\
  \bibnamefont {{Fisher}}},\ }\bibfield  {title} {\enquote {\bibinfo {title}
  {{Majorana zero modes in one-dimensional quantum wires without long-ranged
  superconducting order}},}\ }\href {\doibase 10.1103/PhysRevB.84.195436}
  {\bibfield  {journal} {\bibinfo  {journal} {Phys. Rev. B}\ }\textbf {\bibinfo
  {volume} {84}},\ \bibinfo {eid} {195436} (\bibinfo {year} {2011})},\ \Eprint
  {http://arxiv.org/abs/arXiv:1106.2598} {arXiv:1106.2598} \BibitemShut
  {NoStop}%
\bibitem [{\citenamefont {{Cheng}}\ and\ \citenamefont
  {{Lutchyn}}(2015)}]{Cheng15}%
  \BibitemOpen
  \bibfield  {author} {\bibinfo {author} {\bibfnamefont {Meng}\ \bibnamefont
  {{Cheng}}}\ and\ \bibinfo {author} {\bibfnamefont {Roman}\ \bibnamefont
  {{Lutchyn}}},\ }\bibfield  {title} {\enquote {\bibinfo {title} {{Fractional
  Josephson effect in number-conserving systems}},}\ }\href {\doibase
  10.1103/PhysRevB.92.134516} {\bibfield  {journal} {\bibinfo  {journal} {Phys.
  Rev. B}\ }\textbf {\bibinfo {volume} {92}},\ \bibinfo {eid} {134516}
  (\bibinfo {year} {2015})},\ \Eprint {http://arxiv.org/abs/arXiv:1502.04712}
  {arXiv:1502.04712} \BibitemShut {NoStop}%
\bibitem [{\citenamefont {{Knapp}}\ \emph {et~al.}(2018)\citenamefont
  {{Knapp}}, \citenamefont {{Karzig}}, \citenamefont {{Lutchyn}},\ and\
  \citenamefont {{Nayak}}}]{Knapp17}%
  \BibitemOpen
  \bibfield  {author} {\bibinfo {author} {\bibfnamefont {C.}~\bibnamefont
  {{Knapp}}}, \bibinfo {author} {\bibfnamefont {T.}~\bibnamefont {{Karzig}}},
  \bibinfo {author} {\bibfnamefont {R.~M.}\ \bibnamefont {{Lutchyn}}}, \ and\
  \bibinfo {author} {\bibfnamefont {C.}~\bibnamefont {{Nayak}}},\ }\bibfield
  {title} {\enquote {\bibinfo {title} {{Dephasing of Majorana-based qubits}},}\
  }\href {\doibase 10.1103/PhysRevB.97.125404} {\bibfield  {journal} {\bibinfo
  {journal} {Phys. Rev. B}\ }\textbf {\bibinfo {volume} {97}},\ \bibinfo {eid}
  {125404} (\bibinfo {year} {2018})},\ \Eprint
  {http://arxiv.org/abs/arXiv:1711.03968} {arXiv:1711.03968} \BibitemShut
  {NoStop}%
\bibitem [{\citenamefont {{Snizhko}}\ \emph {et~al.}(2018)\citenamefont
  {{Snizhko}}, \citenamefont {{Egger}},\ and\ \citenamefont
  {{Gefen}}}]{Snizhko18}%
  \BibitemOpen
  \bibfield  {author} {\bibinfo {author} {\bibfnamefont {Kyrylo}\ \bibnamefont
  {{Snizhko}}}, \bibinfo {author} {\bibfnamefont {Reinhold}\ \bibnamefont
  {{Egger}}}, \ and\ \bibinfo {author} {\bibfnamefont {Yuval}\ \bibnamefont
  {{Gefen}}},\ }\bibfield  {title} {\enquote {\bibinfo {title} {{Measurement
  and control of a Coulomb-blockaded parafermion box}},}\ }\href {\doibase
  10.1103/PhysRevB.97.081405} {\bibfield  {journal} {\bibinfo  {journal} {Phys.
  Rev. B}\ }\textbf {\bibinfo {volume} {97}},\ \bibinfo {eid} {081405}
  (\bibinfo {year} {2018})},\ \Eprint {http://arxiv.org/abs/arXiv:1704.03241}
  {arXiv:1704.03241} \BibitemShut {NoStop}%
\bibitem [{\citenamefont {{Seidel}}\ and\ \citenamefont
  {{Lee}}(2005)}]{Seidel05}%
  \BibitemOpen
  \bibfield  {author} {\bibinfo {author} {\bibfnamefont {Alexander}\
  \bibnamefont {{Seidel}}}\ and\ \bibinfo {author} {\bibfnamefont {Dung-Hai}\
  \bibnamefont {{Lee}}},\ }\bibfield  {title} {\enquote {\bibinfo {title} {{The
  Luther-Emery liquid: Spin gap and anomalous flux period}},}\ }\href {\doibase
  10.1103/PhysRevB.71.045113} {\bibfield  {journal} {\bibinfo  {journal} {Phys.
  Rev. B}\ }\textbf {\bibinfo {volume} {71}},\ \bibinfo {eid} {045113}
  (\bibinfo {year} {2005})},\ \Eprint {http://arxiv.org/abs/cond-mat/0402663}
  {cond-mat/0402663} \BibitemShut {NoStop}%
\bibitem [{\citenamefont {Stanescu}\ \emph {et~al.}(2011)\citenamefont
  {Stanescu}, \citenamefont {Lutchyn},\ and\ \citenamefont
  {Das~Sarma}}]{Stanescu2011}%
  \BibitemOpen
  \bibfield  {author} {\bibinfo {author} {\bibfnamefont {Tudor~D.}\
  \bibnamefont {Stanescu}}, \bibinfo {author} {\bibfnamefont {Roman~M.}\
  \bibnamefont {Lutchyn}}, \ and\ \bibinfo {author} {\bibfnamefont
  {S.}~\bibnamefont {Das~Sarma}},\ }\bibfield  {title} {\enquote {\bibinfo
  {title} {Majorana fermions in semiconductor nanowires},}\ }\href {\doibase
  10.1103/PhysRevB.84.144522} {\bibfield  {journal} {\bibinfo  {journal}
  {\prb}\ }\textbf {\bibinfo {volume} {84}},\ \bibinfo {pages} {144522}
  (\bibinfo {year} {2011})}\BibitemShut {NoStop}%
\bibitem [{\citenamefont {{Fidkowski}}\ \emph {et~al.}(2012)\citenamefont
  {{Fidkowski}}, \citenamefont {{Alicea}}, \citenamefont {{Lindner}},
  \citenamefont {{Lutchyn}},\ and\ \citenamefont {{Fisher}}}]{Fidkowski12}%
  \BibitemOpen
  \bibfield  {author} {\bibinfo {author} {\bibfnamefont {Lukasz}\ \bibnamefont
  {{Fidkowski}}}, \bibinfo {author} {\bibfnamefont {Jason}\ \bibnamefont
  {{Alicea}}}, \bibinfo {author} {\bibfnamefont {Netanel~H.}\ \bibnamefont
  {{Lindner}}}, \bibinfo {author} {\bibfnamefont {Roman~M.}\ \bibnamefont
  {{Lutchyn}}}, \ and\ \bibinfo {author} {\bibfnamefont {Matthew P.~A.}\
  \bibnamefont {{Fisher}}},\ }\bibfield  {title} {\enquote {\bibinfo {title}
  {{Universal transport signatures of Majorana fermions in
  superconductor-Luttinger liquid junctions}},}\ }\href {\doibase
  10.1103/PhysRevB.85.245121} {\bibfield  {journal} {\bibinfo  {journal} {Phys.
  Rev. B}\ }\textbf {\bibinfo {volume} {85}},\ \bibinfo {eid} {245121}
  (\bibinfo {year} {2012})},\ \Eprint {http://arxiv.org/abs/arXiv:1203.4818}
  {arXiv:1203.4818} \BibitemShut {NoStop}%
\bibitem [{\citenamefont {{Clarke}}\ \emph {et~al.}(2013)\citenamefont
  {{Clarke}}, \citenamefont {{Alicea}},\ and\ \citenamefont
  {{Shtengel}}}]{Clarke13}%
  \BibitemOpen
  \bibfield  {author} {\bibinfo {author} {\bibfnamefont {David~J.}\
  \bibnamefont {{Clarke}}}, \bibinfo {author} {\bibfnamefont {Jason}\
  \bibnamefont {{Alicea}}}, \ and\ \bibinfo {author} {\bibfnamefont {Kirill}\
  \bibnamefont {{Shtengel}}},\ }\bibfield  {title} {\enquote {\bibinfo {title}
  {{Exotic non-Abelian anyons from conventional fractional quantum Hall
  states}},}\ }\href {\doibase 10.1038/ncomms2340} {\bibfield  {journal}
  {\bibinfo  {journal} {Nature Communications}\ }\textbf {\bibinfo {volume}
  {4}},\ \bibinfo {eid} {1348} (\bibinfo {year} {2013})},\ \Eprint
  {http://arxiv.org/abs/arXiv:1204.5479} {arXiv:1204.5479} \BibitemShut
  {NoStop}%
\bibitem [{\citenamefont {{Fu}}(2010)}]{Fu10}%
  \BibitemOpen
  \bibfield  {author} {\bibinfo {author} {\bibfnamefont {Liang}\ \bibnamefont
  {{Fu}}},\ }\bibfield  {title} {\enquote {\bibinfo {title} {{Electron
  Teleportation via Majorana Bound States in a Mesoscopic Superconductor}},}\
  }\href {\doibase 10.1103/PhysRevLett.104.056402} {\bibfield  {journal}
  {\bibinfo  {journal} {Phys. Rev. Lett}\ }\textbf {\bibinfo {volume} {104}},\
  \bibinfo {eid} {056402} (\bibinfo {year} {2010})},\ \Eprint
  {http://arxiv.org/abs/arXiv:0909.5172} {arXiv:0909.5172} \BibitemShut
  {NoStop}%
\bibitem [{\citenamefont {{van Heck}}\ \emph {et~al.}(2016)\citenamefont {{van
  Heck}}, \citenamefont {{Lutchyn}},\ and\ \citenamefont
  {{Glazman}}}]{vanHeck16}%
  \BibitemOpen
  \bibfield  {author} {\bibinfo {author} {\bibfnamefont {B.}~\bibnamefont {{van
  Heck}}}, \bibinfo {author} {\bibfnamefont {R.~M.}\ \bibnamefont {{Lutchyn}}},
  \ and\ \bibinfo {author} {\bibfnamefont {L.~I.}\ \bibnamefont {{Glazman}}},\
  }\bibfield  {title} {\enquote {\bibinfo {title} {{Conductance of a
  proximitized nanowire in the Coulomb blockade regime}},}\ }\href {\doibase
  10.1103/PhysRevB.93.235431} {\bibfield  {journal} {\bibinfo  {journal} {Phys.
  Rev. B}\ }\textbf {\bibinfo {volume} {93}},\ \bibinfo {eid} {235431}
  (\bibinfo {year} {2016})},\ \Eprint {http://arxiv.org/abs/arXiv:1603.08258}
  {arXiv:1603.08258} \BibitemShut {NoStop}%
\bibitem [{\citenamefont {{Gangadharaiah}}\ \emph {et~al.}(2011)\citenamefont
  {{Gangadharaiah}}, \citenamefont {{Braunecker}}, \citenamefont {{Simon}},\
  and\ \citenamefont {{Loss}}}]{Gangadharaiah11}%
  \BibitemOpen
  \bibfield  {author} {\bibinfo {author} {\bibfnamefont {Suhas}\ \bibnamefont
  {{Gangadharaiah}}}, \bibinfo {author} {\bibfnamefont {Bernd}\ \bibnamefont
  {{Braunecker}}}, \bibinfo {author} {\bibfnamefont {Pascal}\ \bibnamefont
  {{Simon}}}, \ and\ \bibinfo {author} {\bibfnamefont {Daniel}\ \bibnamefont
  {{Loss}}},\ }\bibfield  {title} {\enquote {\bibinfo {title} {{Majorana Edge
  States in Interacting One-Dimensional Systems}},}\ }\href {\doibase
  10.1103/PhysRevLett.107.036801} {\bibfield  {journal} {\bibinfo  {journal}
  {\prl}\ }\textbf {\bibinfo {volume} {107}},\ \bibinfo {eid} {036801}
  (\bibinfo {year} {2011})},\ \Eprint {http://arxiv.org/abs/1101.0094}
  {arXiv:1101.0094 [cond-mat.str-el]} \BibitemShut {NoStop}%
\bibitem [{\citenamefont {{Lobos}}\ \emph {et~al.}(2012)\citenamefont
  {{Lobos}}, \citenamefont {{Lutchyn}},\ and\ \citenamefont {{Das
  Sarma}}}]{Lobos12}%
  \BibitemOpen
  \bibfield  {author} {\bibinfo {author} {\bibfnamefont {Alejandro~M.}\
  \bibnamefont {{Lobos}}}, \bibinfo {author} {\bibfnamefont {Roman~M.}\
  \bibnamefont {{Lutchyn}}}, \ and\ \bibinfo {author} {\bibfnamefont
  {S.}~\bibnamefont {{Das Sarma}}},\ }\bibfield  {title} {\enquote {\bibinfo
  {title} {{Interplay of Disorder and Interaction in Majorana Quantum
  Wires}},}\ }\href {\doibase 10.1103/PhysRevLett.109.146403} {\bibfield
  {journal} {\bibinfo  {journal} {\prl}\ }\textbf {\bibinfo {volume} {109}},\
  \bibinfo {eid} {146403} (\bibinfo {year} {2012})},\ \Eprint
  {http://arxiv.org/abs/1202.2837} {arXiv:1202.2837 [cond-mat.mes-hall]}
  \BibitemShut {NoStop}%
\bibitem [{\citenamefont {Kim}\ \emph {et~al.}(2017)\citenamefont {Kim},
  \citenamefont {Clarke},\ and\ \citenamefont {Lutchyn}}]{Kim}%
  \BibitemOpen
  \bibfield  {author} {\bibinfo {author} {\bibfnamefont {Younghyun}\
  \bibnamefont {Kim}}, \bibinfo {author} {\bibfnamefont {David~J.}\
  \bibnamefont {Clarke}}, \ and\ \bibinfo {author} {\bibfnamefont {Roman~M.}\
  \bibnamefont {Lutchyn}},\ }\bibfield  {title} {\enquote {\bibinfo {title}
  {Coulomb blockade in fractional topological superconductors},}\ }\href
  {\doibase 10.1103/PhysRevB.96.041123} {\bibfield  {journal} {\bibinfo
  {journal} {Phys. Rev. B}\ }\textbf {\bibinfo {volume} {96}},\ \bibinfo
  {pages} {041123} (\bibinfo {year} {2017})}\BibitemShut {NoStop}%
\bibitem [{\citenamefont {{Lutchyn}}\ \emph {et~al.}(2018)\citenamefont
  {{Lutchyn}}, \citenamefont {{Bakkers}}, \citenamefont {{Kouwenhoven}},
  \citenamefont {{Krogstrup}}, \citenamefont {{Marcus}},\ and\ \citenamefont
  {{Oreg}}}]{Lutchyn_review}%
  \BibitemOpen
  \bibfield  {author} {\bibinfo {author} {\bibfnamefont {R.~M.}\ \bibnamefont
  {{Lutchyn}}}, \bibinfo {author} {\bibfnamefont {E.~P.~A.~M.}\ \bibnamefont
  {{Bakkers}}}, \bibinfo {author} {\bibfnamefont {L.~P.}\ \bibnamefont
  {{Kouwenhoven}}}, \bibinfo {author} {\bibfnamefont {P.}~\bibnamefont
  {{Krogstrup}}}, \bibinfo {author} {\bibfnamefont {C.~M.}\ \bibnamefont
  {{Marcus}}}, \ and\ \bibinfo {author} {\bibfnamefont {Y.}~\bibnamefont
  {{Oreg}}},\ }\bibfield  {title} {\enquote {\bibinfo {title} {{Majorana zero
  modes in superconductor-semiconductor heterostructures}},}\ }\href {\doibase
  10.1038/s41578-018-0003-1} {\bibfield  {journal} {\bibinfo  {journal} {Nature
  Reviews Materials}\ }\textbf {\bibinfo {volume} {3}},\ \bibinfo {pages}
  {52--68} (\bibinfo {year} {2018})},\ \Eprint
  {http://arxiv.org/abs/1707.04899} {arXiv:1707.04899 [cond-mat.supr-con]}
  \BibitemShut {NoStop}%
\bibitem [{\citenamefont {Giamarchi}(2004)}]{Giamarchi04}%
  \BibitemOpen
  \bibfield  {author} {\bibinfo {author} {\bibfnamefont {T.}~\bibnamefont
  {Giamarchi}},\ }\href@noop {} {\emph {\bibinfo {title} {Quantum Physics in
  One Dimension}}},\ \bibinfo {edition} {10th}\ ed.\ (\bibinfo  {publisher}
  {Clarendon Press},\ \bibinfo {year} {2004})\BibitemShut {NoStop}%
\end{thebibliography}


\end{document}